\documentclass[sigconf]{acmart}
\usepackage{subcaption}
\usepackage{listings}
\usepackage{booktabs}
\usepackage{enumitem}
\copyrightyear{2026}
\acmYear{2026}
\setcopyright{cc}
\setcctype{by}
\acmConference[CHI '26]{Proceedings of the 2026 CHI Conference on Human Factors in Computing Systems}{April 13--17, 2026}{Barcelona, Spain}
\acmBooktitle{Proceedings of the 2026 CHI Conference on Human Factors in Computing Systems (CHI '26), April 13--17, 2026, Barcelona, Spain}
\acmDOI{w}
\acmISBN{979-8-4007-2278-3/2026/04}

\lstdefinelanguage{Cypher}{
    keywords={MATCH, RETURN, WHERE, CREATE, DELETE, SET, WITH, OPTIONAL, UNWIND, MERGE, FOREACH, DISTINCT, ORDER, BY, ASC, DESC, LIMIT},
    sensitive=true,
    comment=[l]{//},
    morecomment=[s]{/*}{*/},
    string=[b]",
}

\lstdefinestyle{cypherstyle}{
    language=Cypher,
    basicstyle=\ttfamily\footnotesize, 
    keywordstyle=\bfseries,
    commentstyle=\itshape,
    stringstyle=\color{red!70!black},
    numbers=left,
    numberstyle=\tiny,
    stepnumber=1,
    numbersep=5pt,
    frame=single,
    breaklines=true,
    showstringspaces=false,
    tabsize=2
}
\AtBeginDocument{%
  }

\begin{document}

\title{Reflecting on 1,000 Social Media Journeys: \protect\\ Generational Patterns in Platform Transition}

\author{Artur Solomonik}
\orcid{0009-0000-8349-148X}
\email{artur.solomonik@cais-research.de}
\affiliation{%
  \institution{Center for Advanced Internet Studies (CAIS)}
  \city{Bochum}
  \country{Germany}
}

\author{Nicolas Ruiz}
\orcid{0009-0003-8412-9935}
\email{nicolas.ruiz@cais-research.de}
\affiliation{
  \institution{Center for Advanced Internet Studies (CAIS) \\ 
  \& University of Bremen}
  \city{Bochum}
  \country{Germany}
}
\author{Hendrik Heuer}
\orcid{0000-0003-1919-9016}
\email{hendrik.heuer@cais-research.de}
\affiliation{%
  \institution{Center for Advanced Internet Studies (CAIS)}
  \city{Bochum}
  \country{Germany}
}
\affiliation{%
  \institution{University of Wuppertal}
  \city{Wuppertal}
  \country{Germany}
}

\renewcommand{\shortauthors}{Solomonik et al.}

\begin{abstract}
Social media has billions of users, but we still do not fully understand why users prefer one platform over another. Establishing new platforms among already popular competitors is difficult. Prior research has richly documented people's experiences within individual platforms, yet situating those experiences within the entirety of a user's social media experience remains challenging. What platforms have people used, and why have they transitioned between them? We collected data from a quota-based sample of 1,000 U.S. participants. We introduce the concept of \emph{Social Media Journeys} to study the entirety of their social media experiences systematically. We identify push and pull factors across the social media landscape. We also show how different generations adopted social media platforms based on personal needs. With this work, we advance HCI by moving towards holistic perspectives when discussing social media technology, offering new insights for platform design, governance, and regulation.
\end{abstract}



\maketitle
\section{Introduction}

With the recent exodus from Twitter~\cite{HuangMastodon2022}, shifts between social media platforms have gained renewed attention. Such movements are hardly new: users have previously moved from MySpace to Facebook, from Facebook to Instagram, or from Instagram to TikTok. These Journeys are not just one-off Transitions but reveal patterns that reflect how social media use changes across time. Although platforms often follow similar social and technical design principles, the ways they structure content and determine its source remain crucial points of distinction~\cite{zhangFormFrom2024}.  

User migration has been linked to dissatisfaction, attraction to alternatives, and external pressures~\cite{fieslerMoving2020,quelle2025academicsleavingtwitterbluesky}. Such dynamics intersect with ongoing regulatory debates: for example, Australia’s ban on social media under the age of 16~\cite{Australia2025} and the UK’s Online Safety Act, which requires companies to implement age assurance measures~\cite{Guardian_OnlineSafetyAct_FreeSpeech_2025}. These measures illustrate the high stakes of contemporary social media governance, where questions of youth safety, privacy, and free expression remain contested. While our study does not directly evaluate these policies, as the assessment of children’s well-being in relation to social media use remains ongoing and complex~\cite{kross2021social, kim2017impact, richards2015impact}, it takes inspiration from this moment of change to ask how social media might be designed and governed in ways that better align with people’s needs. To inform these complex processes of design and governance, we introduce a new approach to understanding who uses which platform and why. While sources such as the Reuters Digital News Report provide robust demographic breakdowns of platform use~\cite {reuters2025}, they primarily describe platform adoption at a single point in time. As a result, they cannot show how people transition between platforms, what prompts these changes, or how earlier experiences shape later choices. This reliance on static demographic categories risks overgeneralization, especially when individual histories involve multiple overlapping platforms and diverse experiences that can't be measured by time spent on a platform alone~\cite{Jungselius2018conceptualizing}. These limitations highlight the need for approaches that capture the temporal and relational dimensions of social media use. As numerous platforms that once shaped a person's internet experiences are now defunct, abandoned, or terminated, our work aims to address the empirical gaps through a more holistic and systematic retrospection alongside existing longitudinal work~\cite{jungselius2025long} and reported perspectives on intersections of distinct platforms and age groups~\cite{boyd2014itscomplicated,  schlesingerSituatedAnonymityImpacts2017,boczkowski2018multiple, taber2023norms}. Our work extends this literature by tracing how people move across platforms across their lives, giving policymakers, researchers, and designers a holistic view of social media use as a sequence of connected shifts rather than separate points in time.

To capture how people move across platforms, we introduce the concept of the \textit{Social Media Journey}. A \textit{Journey} represents a series of directed movements of an individual’s activity across the social media landscapes. Each of these movements, which we call \textit{Transitions}, marks the shift from one platform to another for particular reasons. A Transition, in this sense, is not a single act of joining or leaving a specific social media platform, but rather a contextually situated movement between platforms that reveals how users adapt to the changing social media landscape. We chose the term transition as an alternative to social media migration~\cite{fieslerMoving2020, solomonikJourneys2025}. Concepts such as platform-swinging~\cite{polyplatformContext2019}, context collapse~\cite{boyd2014itscomplicated}, and the digital footprint informed our decision to introduce a more general term to capture the movement between platforms. Moreover, we use Transitions as an analytical lens to make sense of sequences of platform adoption by situating them within the push and pull factors that draw people toward or away from particular platforms, and how these factors intersect with design affordances, community practices, and personal needs. We aim to provide conceptual and empirical tools that help both scholars and practitioners better understand how platforms succeed, or become meaningful in people’s lives, and ultimately to inform the creation and governance of more inclusive and resilient social media systems.

Social Media Journeys are not simply personal choices but reflections of how design, community, and context interact across the social media landscape. Decisions with broad-ranging consequences about design, governance, or regulation require strong empirical grounding in these lived experiences. To this end, we ask: 
\begin{enumerate}[label=\textbf{RQ\arabic*:}, leftmargin=*, align=left]
  \item \textbf{What roles do platforms play within self-reported social media journeys?}
  \item \textbf{What do social media journeys reveal about people’s reasons for using social media?}
  \item \textbf{How have platform design features and social factors influenced different generations throughout their journeys?}
\end{enumerate}

To answer these questions, we conducted a mixed-methods study with a quota-based sample of 1000 U.S. citizens (492 women, 12 non-binary) that was matched to the U.S. Census and stratified according to the U.S. income distribution. Participants created visual representations of their \textit{Social Media Journeys} using a custom-built online tool. The tool allowed participants to connect platforms with arrows, annotate reasons for transitioning, and add free-text details, producing graph structures that reflect both movement and motivation. This design enabled us to capture Journeys not just as isolated accounts, but as complex networks of people, platforms, and Transitions.  

We then analyzed these graphs both quantitatively and qualitatively. Quantitatively, the graph representation allowed us to compute measures such as in-degree and out-degree for platforms, highlighting where activity accumulated or declined. Qualitatively, participants’ reasons for moving situated the directions within the graph in lived experience. In combining both lenses, we foreground not just where people go, but why, and what this reveals about the shifting landscape of social media.

We provide policymakers and future social media designers with insights into how people’s experiences with social media, together with the relationships between competing platforms, shape their present-day use. Specifically, we identify the roles that platforms occupy within users’ Journeys, and we highlight the push and pull dynamics that make platforms functionally and socially complementary rather than interchangeable. By comparing across generations, we show how social media is taken up differently over time.  

Our study shows that Transitions between platforms are not random but responses to three factors. Design features reveal why some platforms persist while others fade. Shifts within communities show how users exercise agency in the face of algorithmic feeds or policy changes. Broader societal changes highlight the trade-offs people make between visibility, privacy, and community. By situating these dynamics in a generational perspective, we provide a grounded understanding of Social Media Journeys that can guide platform designers in creating more responsive features, policymakers in regulating more effectively, and researchers in developing contextual approaches to studying social media.
\section{Background}
\label{sec:background}
Social media has evolved from isolated communication tools into complex, interconnected ecosystems that shape how people interact, express themselves, and build communities. Understanding these systems requires frameworks that account not only for technological features but also for users’ lived experiences, motivations, and adaptations over time. This background reviews existing research on platform typologies, digital migration, longitudinal and generational perspectives, and social media appropriation. Taken together, these four areas of prior work illustrate how platform structures, migration dynamics, temporality, and lived experience intersect, but they have not yet been integrated into a single holistic framework that captures multi-platform, longitudinal, and generational movement.

\subsection{Platform Taxonomies and the Form--From Framework}
\label{sec:taxonomies}
Analyzing Social Media Journeys requires a definition that can compare systems while capturing lived experience. Ouirdi et al. classify platforms by users, content formats, and functions, emphasizing strategies for attracting and retaining members~\cite{ouirdiSocialMediaConceptualization2014}. DeVito extends this view by incorporating users’ folk theories~\cite{devito2017algorithms}, showing how perceptions of curation shape behavior and platform evolution~\cite{devitoAdapt2021}. Santos further positions social media as a political and economic actor~\cite {santosSocalledUGCUpdated2022}, underscoring its societal impact beyond content and community alone. While prior work shows that social media platforms are dynamic, shaped by designers and audiences, there is no holistic way to measure why people continually navigate, adopt, and abandon them~\cite{hou2022staying,polyplatformContext2019}. 

We build on the \textit{Form--From} framework, which distinguishes platforms along two dimensions: the \textit{Form} of content (flat vs. threaded) and the \textit{From} of content delivery (spaces, networks, or the commons). According to Form--From, \textit{Flat} platforms are those where posts are self-contained without direct links to other posts, whereas \textit{threaded} platforms connect posts through replies or comments. The model also differentiates between \textit{Spaces}, \textit{Networks}, and \textit{Commons} based on how content is shared. In \textit{Spaces}, users must actively seek out content within distinct areas such as channels or groups. In \textit{Networks}, content is shared with connected accounts, like followers or friends. In \textit{Commons}, content is shared platform-wide and distributed algorithmically, as seen with features like a ``For You'' page. This 2x3 model yields six classifications: \textit{Flat Spaces} (e.g., WhatsApp), \textit{Threaded Spaces} (e.g., Early Twitter), \textit{Flat Network} (e.g., Early Facebook), \textit{Threaded Network} (e.g., Instagram), \textit{Flat Commons} (e.g., Pinterest), and \textit{Threaded Commons} (e.g., Reddit)~\cite{zhangFormFrom2024}. Platforms can blend features, but are typically designed around one combination, shaping content encounter and interaction~\cite{meyer2020socialaffordances}. This lens allows us to see not only that users moved ``from Twitter to TikTok,'' but also from a \textit{Flat Network} to a \textit{Flat Commons}, producing generalizable knowledge for HCI.

\subsection{Push--Pull Dynamics and Digital Migration}
\label{sec:pushpull}

Chang et al. describe a push--pull--mooring framework for migration, where dissatisfaction (push), attraction to alternatives (pull), and switching costs (mooring) jointly shape decisions~\cite{changPush2013,lee1966theory}. Prior research on distinct migration cases emphasize that non-use is often partial, involving reconfiguration of platform-specific social relations~\cite{edwards2021migration, baumernonuse, hughes2012tale}. High-profile cases, such as the leadership change at Twitter, illustrate these dynamics: academics often moved only after their networks were established on Bluesky~\cite{quelle2025academicsleavingtwitterbluesky}. Similar dynamics have been observed during earlier platform crises, where network persistence often delayed large-scale migration~\cite{gillespie2018custodians}. Protests on Reddit regarding the platform's policy on third-party applications revealed how platform moderators how despite people leaving, the amount of hate speech decreased drastically~\cite{chandrasekharan2017stay, marias2016reddit}. Beyond identity-specific practices, appropriation also shapes broader user communities.

Migration is not purely structural but also experiential. Kross et al. show that design and exposure shape emotional outcomes~\cite{kross2021social}, while Fiesler and Dym demonstrate fandom migrations in response to policy, design, and norms~\cite{fieslerMoving2020, pearce2011communities, brubaker2016departing, plantin2018infrastructure}. Jacobs et al. trace long-term network assembly patterns, revealing heterogeneous cross-platform movement~\cite{jacobs2015assembling}. These studies show that migration is negotiated and interconnected; however, most migration research focuses on isolated cases or on a single cohort, leaving cross-generational, multi-platform patterns underexplored.

Prior work often describes migration as movement toward a single, specific use case~\cite{fieslerMoving2020, solomonikJourneys2025}. However, people’s behaviors on the internet are more complex. Context collapse refers to situations where people must deal with different social groups at the same time, each with its own expectations, which can make it difficult to decide how to act or communicate~\cite{boyd2014itscomplicated}. Platform swinging describes how users move back and forth between platforms as their needs or the platforms’ features change~\cite{polyplatformContext2019}. Transitions capture these dynamics and therefore go beyond linear descriptions of migration to address the more overlapping and simultaneous ways people adopt and use platforms.

\subsection{Longitudinal and Generational Perspectives on Social Media Adoption}
\label{sec:longitudinal}

Longitudinal studies are essential for understanding temporal change and context-specific effects. Foundational work emphasizes tracking user behavior over time to capture evolving practices alongside technological affordances~\cite{Zhao2013manyfaces, Jungselius2018conceptualizing, Bernstein2013quantifying}. Reviews of longitudinal HCI research highlight their value for methodological rigor~\cite{Kjaerup2021longitudinal, Audulv2023timeandchange, Bayer2020socialmediaelements, bolton2013understanding}. The ten-year longitudinal study by Jungselius and Weilenmann traces how perceptions of social media shifted along three trajectories at three points in time: from public to private interaction, from producing to consuming, and from fun to problematic~\cite{jungselius2025long}. Our study aligns with this tradition but addresses its limits. We connect retrospective recall to intergenerational comparisons, situating personal experiences within broader historical trends in the social media landscape.
Generational analysis can risk overgeneralization; we follow guidance from the Pew Research Center, which argues that such comparisons are appropriate when historical data allow us to examine different generations at similar stages of life~\cite{parker2023pew}. This position provides a conceptual foundation for engaging with generational perspectives without reifying them. For our work, these labels align with research that informs policymakers, such as the Reuters Digital News Report~\cite{reuters2025} and provides generalizable demographics of social media use. The long timeline of platforms emerging and fading away at different moments means that age groups encountered them at varying stages of their lives. It provides a useful historical grounding for future generational comparison.

Boczkowski et al. illustrate how young users in Argentina manage repertoires of multiple platforms, attaching distinct meanings to each (e.g., WhatsApp as a versatile communication tool, Facebook as a socially acceptable means of self-presentation, Instagram as a stylized display)~\cite{boczkowski2018multiple}. Taber et al. extend this focus to Millennials and Gen-Z, showing that posting decisions are guided less by affordances than by alignment with platform-specific social norms~\cite{taber2023norms}. These studies underline the importance of temporal and generational perspectives, as well as the social construction of platform roles~\cite{boyd2014itscomplicated}. 
Empirical work has further shown that generational differences manifest within platforms: teens engage in highly self-representational practices on Instagram, make strategic use of tagging culture, and actively curate their posts through deletion, whereas adults tend to post more diverse content and place less emphasis on performative metrics~\cite{jang2015generationLike}.
At the opposite end of the spectrum, research with older adults demonstrates that non-use is often an active preference rather than a capability gap, shaped by concerns about privacy, time, irrelevance, and the erosion of meaningful communication~\cite{hope2014oldCommunicationDigital}.
Recent HCI work conceptualizes ageing as a multidimensional phenomenon rather than a single demographic marker~\cite{knowles2025Trust}, a perspective that aligns with evidence that trust in social media platforms and interventions tends to decrease with age, indicating distinctive orientations toward risk, credibility, and governance.

We extend this work with a method that captures retrospective, multi-platform experiences across generations. By representing adoption and reconfiguration in a single structure, Social Media Journeys reveal how design features, social norms, and personal histories interact over time.
This framing allows us to engage with generational patterns while maintaining the caution advised by Parker: generational lenses are most analytically defensible when situated within longitudinal evidence~\cite{parker2023pew}. Because our tool traces self-reported platforms from multiple decades--including defunct, dominant, and transitional services--they provide the historical grounding necessary for responsible generational interpretation. Our retrospective, cross-platform approach shows how people transition between platforms in relation to broader social and technological shifts while using generational categories as a historically comparative lens.

\subsection{Social Media Appropriation and Lived Experience}
\label{sec:appropriation}

Social media appropriation refers to how users adapt platforms to fit their needs, treating systems as flexible tools rather than fixed artefacts~\cite{dourish2003appropriation}. This is supported by studies showing that people of all ages adapt platforms to their own needs, use several platforms at once, and manage challenges beyond identity-specific concerns~\cite{predictorsSocialMediaUse2021, polyplatformContext2019}. This process reflects the situated, dynamic, and sometimes subversive ways users enact control, create ownership, and respond to changing social and technical environments. Appropriation encompasses integration into broader social, identity, and personal practices~\cite{light2018social}.

Haimson conceptualizes social media as social transition machinery, showing how trans people distributed aspects of self across multiple sites during gender transition, with platforms collectively scaffolding identity change~\cite{haimsonTransitionMachinery2018}. Building on Marwick and boyd’s notion of context collapse, it is impossible to anticipate one's audience, leaving it to the user's imagination~\cite{marwick2011tweet}. DeVito et al. describe LGBTQ+ users’ personal social media ecosystems, managing self-presentation across overlapping platforms to balance visibility, safety, and expression~\cite{deVitoTooGay2018}. Haimson et al. describe Tumblr as a trans technology, uniquely enabling experimentation, temporality, and intersectional community-building~\cite{haimsonTumblrWasTrans2021}. Buss et al. show trans users strategically curate identity across networks, tailoring feeds for affirmation~\cite{buss2022transgender}. Kender and Spiel emphasize the value of flexible, anti-normative design in online spaces for supporting marginalized communities. In their work on autistic social media, they argue for customization, non-prescriptiveness, and interest-centric, granular publics to enable thriving neurodiverse communities~\cite{kenderBanalAutistic2023}. In their study with trans participants, they highlight how experiences of algorithmic and structural violence reveal the limitations of large, profile-centric platforms and motivate the creation of interest- and community-oriented spaces that balance safety, freedom, and meaningful interaction~\cite{kender2025machine}.

These studies establish the experiential qualities of online spaces from marginalized perspectives. Using Social Media Journeys, we extend understanding beyond high-stakes identity management to generationally distinct strategies of moderation, selective participation, and well-being-oriented curation.
\section{Methods: Graph-Based Journey Analysis}
\label{sec:methods}
\begin{figure*}[ht]
    \centering
    \includegraphics[width=1.5\columnwidth]{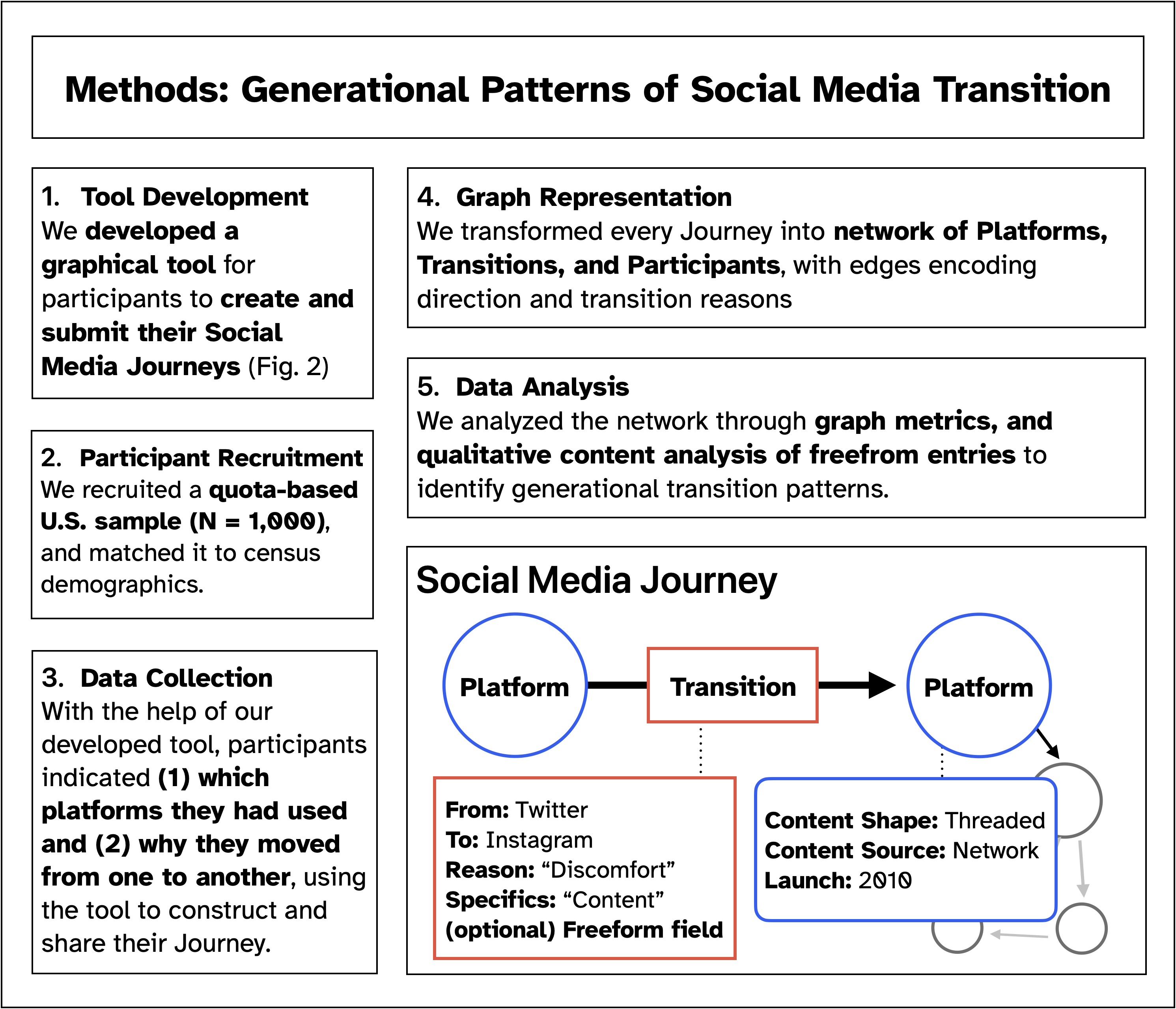}
    \caption{\textbf{Study Procedure:} Our participants used our tool to create their Social Media Journey, which we analyzed through graph metrics and qualitative content analysis.}
    \label{fig:study-procedure}
    \Description{A methods overview showing tool development, recruitment, data collection, graph representation, and analysis, plus an example illustrating a platform-to-platform transition.}
\end{figure*}

To investigate how people transition between social media platforms, we developed a graph-based data collection tool and performed a mixed-methods analysis. In this section, we describe the design of our tool, the procedures for collecting and structuring participant-generated graphs, and the analytic steps used to integrate quantitative graph metrics with qualitative free text entries. Figure~\ref{fig:study-procedure} illustrates our study procedure. We first outline the tool interface and data collection process, followed by details of our participant sample. We then explain how we represented the resulting Journeys as graph structures and how we queried them using Cypher, a declarative querying language similar to SQL~\cite{holzschuher2013performance}. Finally, we describe our analytic approach for interpreting migration patterns, Transition reasons, and generational differences.

\subsection{Tool Design and Data Collection Interface}
\label{sec:methods-tool}
\begin{figure*}
    \includegraphics[width=\linewidth]{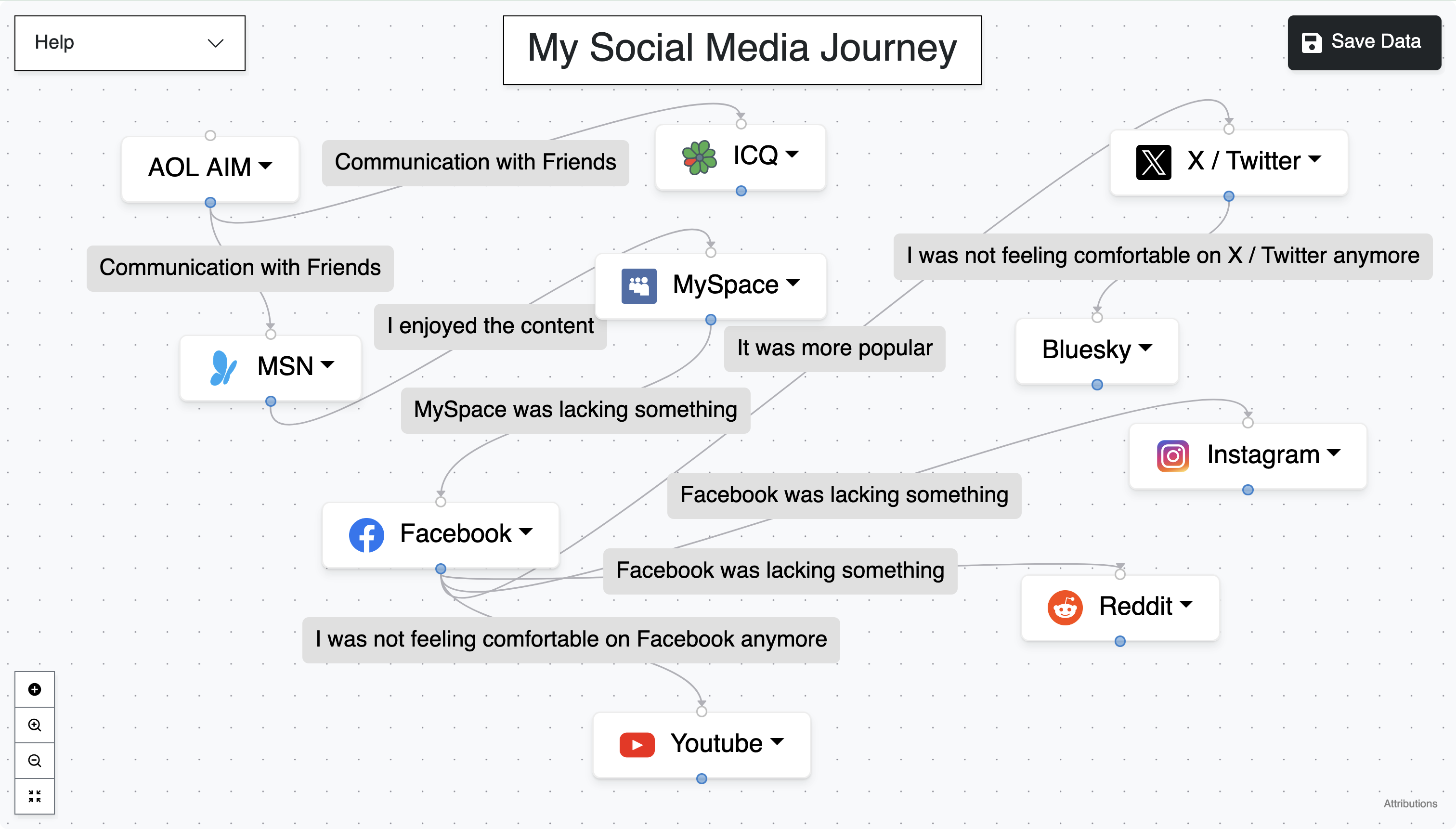}
    \caption{Tool interface for creating a participant's Social Media Journey. To capture what platforms people used (RQ1) and why (RQ2), participants self-reported their Social Media Journey as a graph structure.}
    \Description{Tool interface used for creating a participant’s Social Media Journey, where platforms and reasons for transitions are represented as a graph. Each platform is depicted as a node, including AOL AIM, MSN, ICQ, MySpace, Facebook, Twitter, Instagram, Reddit, YouTube, and Bluesky. Arrows between platforms represent Transitions, with labels describing participants’ reasons for moving. Early platforms such as AOL AIM, MSN, and ICQ are associated with communication with friends. MySpace attracted participants because they enjoyed the content and found it more popular, but it was also described as lacking something, leading to a Transition toward Facebook. Facebook similarly was seen as lacking features or becoming uncomfortable, which prompted moves to Instagram, Reddit, and YouTube. Twitter shows a similar pattern, where participants reported leaving because they no longer felt comfortable, with some transitioning to Bluesky. Overall, the figure highlights how journeys are shaped by both the appeal of new platforms and dissatisfaction with existing ones.}
\end{figure*}
\label{fig:tool}

We introduce the concept of a Journey as a theoretical construct for which we have developed a dedicated data collection tool. Under the supervision of the first author, the second author developed the graphical interface for creating a Social Media Journey shown in Figure~\ref{fig:tool}~\footnote{The source code and deployment instructions are provided on a public GitHub repository~\cite{social-media-journeys}}. A Journey can be created by adding platforms to the workspace and connecting them with a line. When connecting two platforms, a window pops up to ask for the reason for moving from one platform to the other.

Our tool features a default selection of platforms sourced from the Reuters Digital News Report~\cite{reuters2025} while allowing custom free text entries by our participants. The list of options was sourced from prior qualitative findings on social media migration~\cite{fieslerMoving2020,solomonikJourneys2025}. In particular, we compared both papers’ distinction between Platform-based/Community-based Reasons~\cite{fieslerMoving2020} and Platform/Environmental Motivators~\cite{solomonikJourneys2025}, and formulated which reasons could be selected for moving between platforms: “It was more popular”, “Communication with family”, “I enjoyed the content”, “The platform seemed more secure than WhatsApp”, and “I was not feeling comfortable on X/Twitter anymore”. We further categorized these reasons to guide participants through the selection process. General reasons (i.e., “appeal of feature”) revealed sub-options (e.g., messaging, feeds, posting), and participants could also add free-text comments to describe additional details of their Journeys.

\subsection{Participant Recruitment}
\label{sec:methods-participants}
Our quota-based sample included 1,000 participants from the United States recruited using a professional market research panel~\cite{hartman2023connect}. The sample was matched to the U.S. Census~\cite{malik2020determinants} and was stratified to the U.S. income distribution~\cite{DAVIDAI2018138}. Following the HCI guidelines for gender equity and inclusivity by Scheuerman et al.~\cite{scheuermanGender2020}, we found that the US Census does not yet represent non-binary people. To ensure to include their experiences, we oversampled them by 1.6\% following the findings of Brown (2022), who estimates the share of Americans whose gender is different from the sex they were assigned at birth~\cite{brownNb2022}. Oversampling meant including 16 non-binary people in the panel's census-matched sampling criteria of which 12 participated in the study. The participants' mean age was 44.6 years (SD = 15.2). The youngest participant was 18 and the oldest was 79. Respondents were paid \$2.95 for the survey, which took 10 minutes on average.

\subsection{Data Collection}
\label{sec:methods-data}
Our participants were asked to create their Social Media Journey with the help of the following questions: ``What platforms have you used in the past up until now? Why have you moved from one platform to the other? Moving to a new platform does not necessarily mean abandoning the other.'' Participants added platforms they have used from a selection sourced from the most frequently reported platforms from the Reuters Digital News report~\cite{reuters2025}, as well as their own free-text inputs. They added connections between these platforms, where they described the reasons for their Transition as described in Section~\ref{sec:methods-tool}. After validating their participation, participants were compensated for their work through the market research panel. We screened the data and cleaned it for further processing in a graph database system.

\subsection{Graph-Based Data Representation}
\label{sec:methods-graph}
As participants expressed their Social Media Journeys as networks of platforms and directional transitions, representing the data as a graph structure enabled us to systematically compare all participants. We exported and represented the data as a network graph with nodes representing platforms, and connections representing the Transition reasons. We used the graph query language Cypher to query our data, as it allowed us to filter the data according to specific criteria explicitly. 

\begin{figure*}[ht]
\centering
\begin{minipage}[t]{0.48\textwidth}
\begin{lstlisting}[style=cypherstyle, caption={Cypher query for counting the incoming Transitions (in-degree) for a platform},captionpos=t]
MATCH (m:Transition)-[:TO]->(p:Platform)
RETURN p.name AS Platform, COUNT(m) AS IncomingTransitions
ORDER BY IncomingTransitions DESC
\end{lstlisting}
\label{lst:cypher-indegree}
\end{minipage}\hfill
\begin{minipage}[t]{0.48\textwidth}
\begin{lstlisting}[style=cypherstyle, caption={Cypher query for counting Transition reasons when transitioning from MySpace to Facebook},captionpos=t]
MATCH (p:Participant)-[:PERFORMED]->(m:Transition)
MATCH (m)-[:FROM]->(a:Platform{Name: 'MySpace'})
MATCH (m)-[:TO]->(b:Platform{Name: 'Facebook'})
RETURN 
  m.Reason as Reason, count(*) as ReasonCount
\end{lstlisting}
\label{lst:cypher-transition}
\end{minipage}

\Description{Two listings depicting example Cypher queries. The left listing presents a Cypher query designed to calculate the number of incoming Transitions, or in-degree, for each platform in the dataset. It matches Transition events that point to a platform node, retrieves the platform’s name along with a count of these Transitions, and then orders the results in descending order based on the number of incoming Transitions. This allows identification of the platforms that participants most frequently moved to. The right listing shows a Cypher query used to count the reasons participants gave for transitioning specifically from MySpace to Facebook. It first matches participants who performed a Transition, filters those Transitions starting from MySpace and ending at Facebook, and then returns the Transition reasons along with the number of times each reason was reported. This highlights the most common motivations for transitioning between the two platforms.}
\end{figure*}
\label{fig:cypher-examples}
Figure~\ref{fig:cypher-examples} shows what examples for a Cypher query looked like. For every platform, we computed in-degree and out-degree~(RQ1). The in-degree represents how many participants have shifted their activity towards a platform, while the out-degree shows activity shifts away from it. For every platform mentioned, we annotated its content shape (Flat, or Threaded), and content source (Spaces, Network, or Commons) according to the Form-From framework~\cite{zhangFormFrom2024}. We also added the launch date of the platform to follow the historical findings on the evolution of design spaces by Zhang et al. Reported Transition reasons were coded after the distinction of push and pull factors of social media migration by Chang et al.~\cite{changPush2013}. From the perspective of the platform transitioned to, a pulling factor is attributed to desirable characteristics (i.e., Enjoyable Content, Popularity with Friends), while pushing factors are attributed to undesirable characteristics (i.e., Discomfort, Service Termination). As mooring factors are Transition constraints (i.e., sunk costs, adoption costs)~\cite{JUNG2017139}, we could not deduce them from the self-reported data reliably. 

\subsection{Analysis}
\label{sec:methods-analysis}
The first author conducted, what we call, a Graph-Based Mixed-Methods Analysis to synthesize the quantitative values of graph-based data with the qualitative free-text responses gathered by our tool. This synthesis allowed us to interpret the quantitative findings in light of participants’ lived experiences.

We began by categorizing all platforms according to the Form–From Framework by Zhang et al.~\cite{zhangFormFrom2024}. For platforms not included in their original analysis, categories were assigned following their methodological guidelines; unavailable platforms were additionally examined using the Internet Archive’s Wayback Machine~\cite{waybackmachine} by examining the first entry of the most-frequently archived date.

Next, we assigned push- and pull dynamics to all reported reasons following the model by Chang et al.~\cite{changPush2013}. To ensure comparability across participants while still capturing nuanced or unexpected motivations, we combined predefined reason categories with optional free-text input. On average, participants wrote 49.01 characters (std = 38.93), the minimum was 7, and the maximum was 316. All free-text entries were then analyzed using qualitative content analysis~\cite{mayringQualitativeContentAnalysis20045}, a method similar to thematic analysis~\cite{braunThematicAnalysis2006}. Based on axial coding principles~\cite{corbinGroundedTheory2014}, the first author coded all free.text responses while considering platform categorization and participant generation. Codes were refined and clustered in weekly meetings with the third author, resulting in synthesized findings for each generation of Social Media Journeys.

For the quantitative analysis, we counted all Transitions and distributed them across the different reported reasons (RQ2). As we were interested in generalizable insights for Social Medie Journeys among different age groups, we categorized every age by generations according to the Reuters Digital News Report~\cite{reuters2025} (RQ3). To understand the roles of individual platforms within these Journeys, we examined platform properties and counted the most frequently-reported platforms, as well as each platform’s inward and outward connections. These counts enabled more complex queries, such as listing the reasons participants reported for using Facebook when transitioning to MySpace.

\section{Results}
\begin{enumerate}[label=\textbf{RQ\arabic*:}, leftmargin=*, align=left]
  \item \textbf{What roles do platforms play within self-reported social media journeys?}
  \item \textbf{What do social media journeys reveal about people’s reasons for using social media?}
  \item \textbf{How have platform design features and social factors influenced different generations throughout their journeys?}
\end{enumerate}

\subsection{What roles do platforms play within self-reported Social Media Journeys? (RQ1)}
\label{sec:rq1}
\begin{figure}[ht]
    \centering
    \includegraphics[width=\columnwidth]{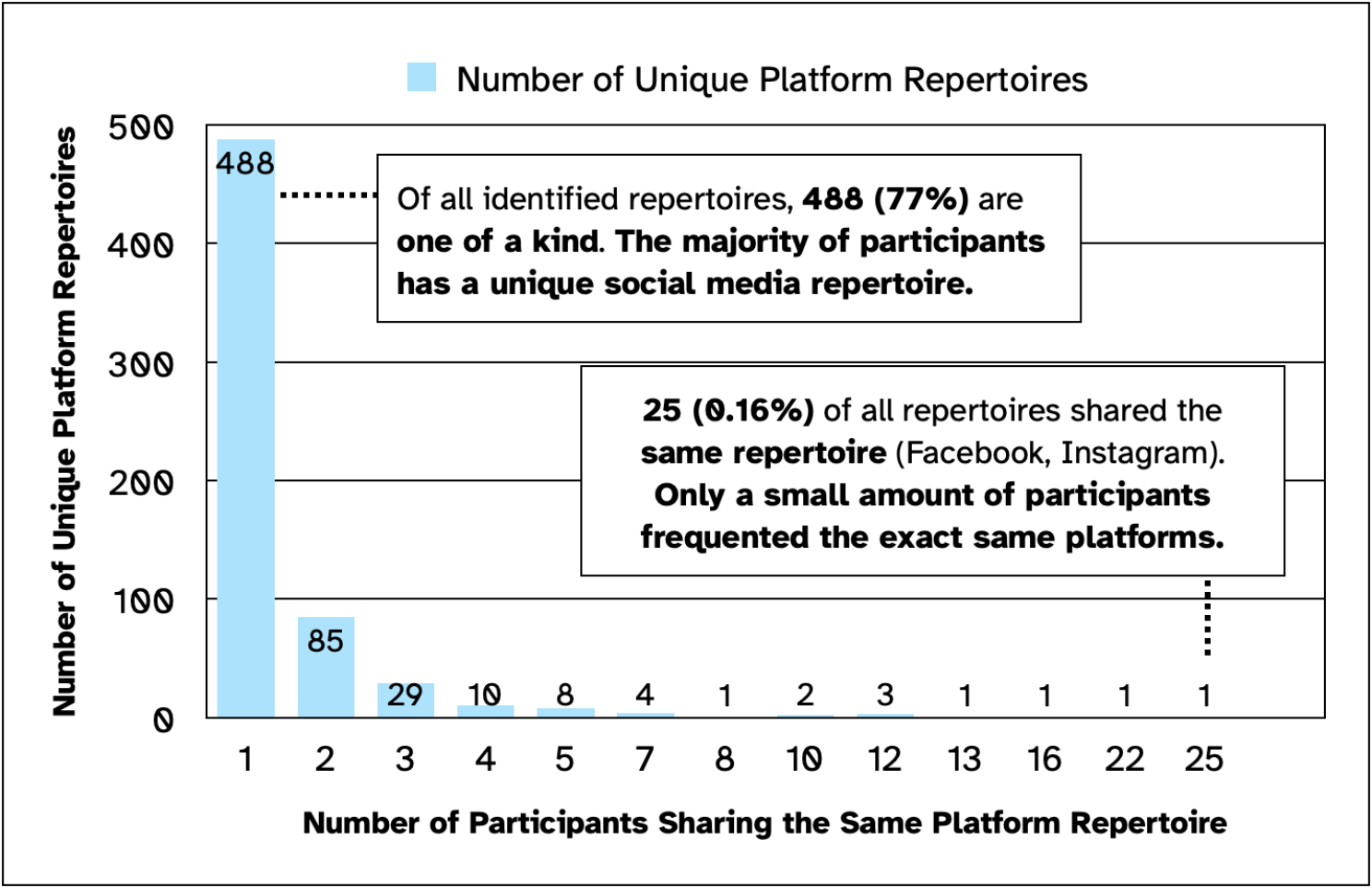}
    \caption{Long-tail distribution of platform adoption: most identified platform repertoires are unique, with only a small share of participants sharing the same repertoire.}
    \label{fig:long-tail}
    \Description{A long-tail bar chart showing that most participants have unique social media platform repertoires, with 488 one-of-a-kind cases and very few shared repertoires.}
\end{figure}

To answer this question, we examined the repertoire of platforms that participants reported as part of their Journeys. The 1,000 participants mentioned 76 unique platforms. On average, participants listed 5.3 platforms~(median = 5, SD = 2.4). For the analysis, we observed the 18 most frequent platforms.

For every participant, we defined their platform repertoire as all platforms present in their Journey. Then, we counted how many participants had the same platform repertoire: Figure~\ref{fig:long-tail} reveals a long-tail distribution of how many people adopted the same platforms. The most common platform repertoire had two platforms: \texttt{Facebook}, \texttt{Instagram}~(2.5\% of all participants). With other repertoires such as \texttt{MySpace}, \texttt{Facebook}, \texttt{Instagram}~(2.1\% of all participants), and \texttt{Facebook}, \texttt{FB Messenger}, \texttt{Instagram}, \texttt{LinkedIn}~(1.6\% of all participants), the long-tail head consists of only 13.2\% of all Journeys. 48.5\% of Journeys are unique, appearing only once in the entire sample.

This finding shows that there is no single dominant Journey through the social media landscape. Instead, Journeys are mostly fragmented and highly individualized, which is difficult to capture in a survey format such as the Reuters Digital News Report~\cite{reuters2025}. The few recurring patterns hint at shared experiences, but the long tail emphasizes that platform adoption is shaped by personal choices and circumstances. This uniqueness informs our later analysis of generational differences: even if certain platforms are mainly used by one generation, the Journeys themselves reveal more nuanced experiences that go beyond simple patterns of adoption.

\subsection{What do Social Media Journeys reveal about people’s reasons for using social media? (RQ2)}
\label{sec:rq2}
To understand people’s reasons for using social media, we examined 3,533 Transition reasons across our quota-based sample. The most frequent Transitions were \texttt{Facebook} $\rightarrow$ \texttt{Instagram}(9.5\%), \texttt{MySpace} $\rightarrow$ \texttt{Facebook}(8.6\%) and \texttt{Facebook} $\rightarrow$ \texttt{FB Messenger}(5.7\%). Every fifth Transition was motivated by the platform’s popularity, which was evenly split: about half mentioned the platform was popular among the general public, and the other half noted it was popular among their friends. In comparison, 7.1\% of participants transitioned to communicating with friends, and 3.4\% to communicating with family. 

\begin{figure*}[t]
  \centering
  \begin{subfigure}[t]{0.49\textwidth}

  \includegraphics[width=0.95\columnwidth]{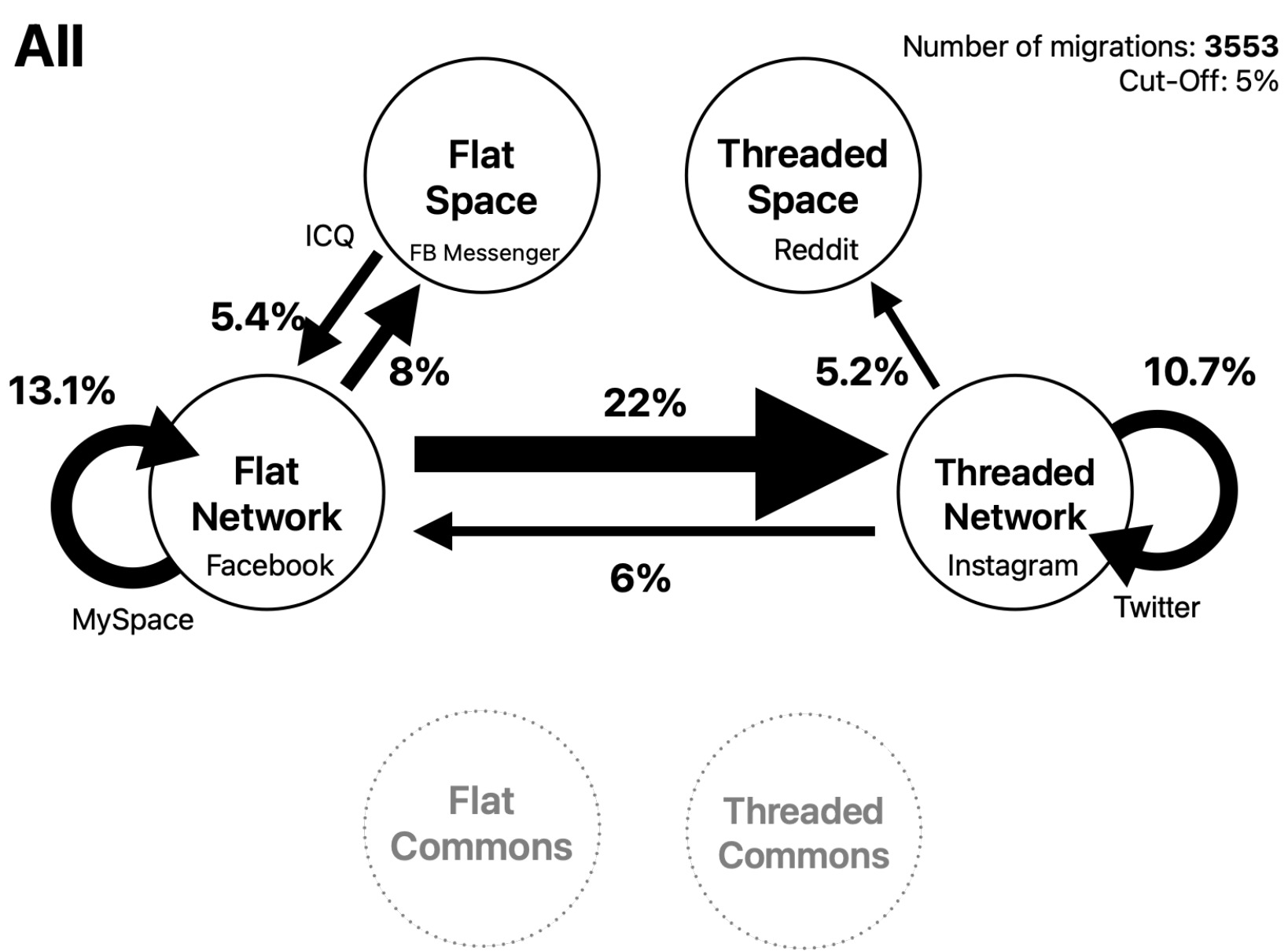}
  \caption{Cross-generational overview of Social Media Journeys across the Form--From design space. Each arrow represents a flow of Journeys between categories, with thickness scaled by frequency. Annotations show the most frequently mentioned platform for each Journey. This visualization highlights shared trajectories as well as generational differences.}
  \label{fig:crossgen}
  \Description{Transition flows between major platform categories, based on 3,553 Transitions with a 5\% cut-off applied. Circles represent categories: Flat Network (Facebook, MySpace), Threaded Network (Instagram, Twitter), Flat Space (Messenger, ICQ), and Threaded Space (Reddit). Arrows between circles indicate Transition directions, with percentages showing flow size. The largest flow is from Facebook to Instagram at 22\%, with 6\% moving in the opposite direction. MySpace shows a 13.1\% self-loop, while Instagram (10.7\%) and Twitter (10.7\%) also retain significant shares. Smaller flows include 8\% from Facebook to Messenger, 5.4\% from Facebook to ICQ, and 5.2\% from Instagram to Reddit. The visualization emphasizes the strongest bidirectional exchange between Facebook and Instagram, alongside notable retention within individual platforms.}
  \end{subfigure}
  \hfill
  \begin{subfigure}[t]{0.49\textwidth}
    \includegraphics[width=\columnwidth]{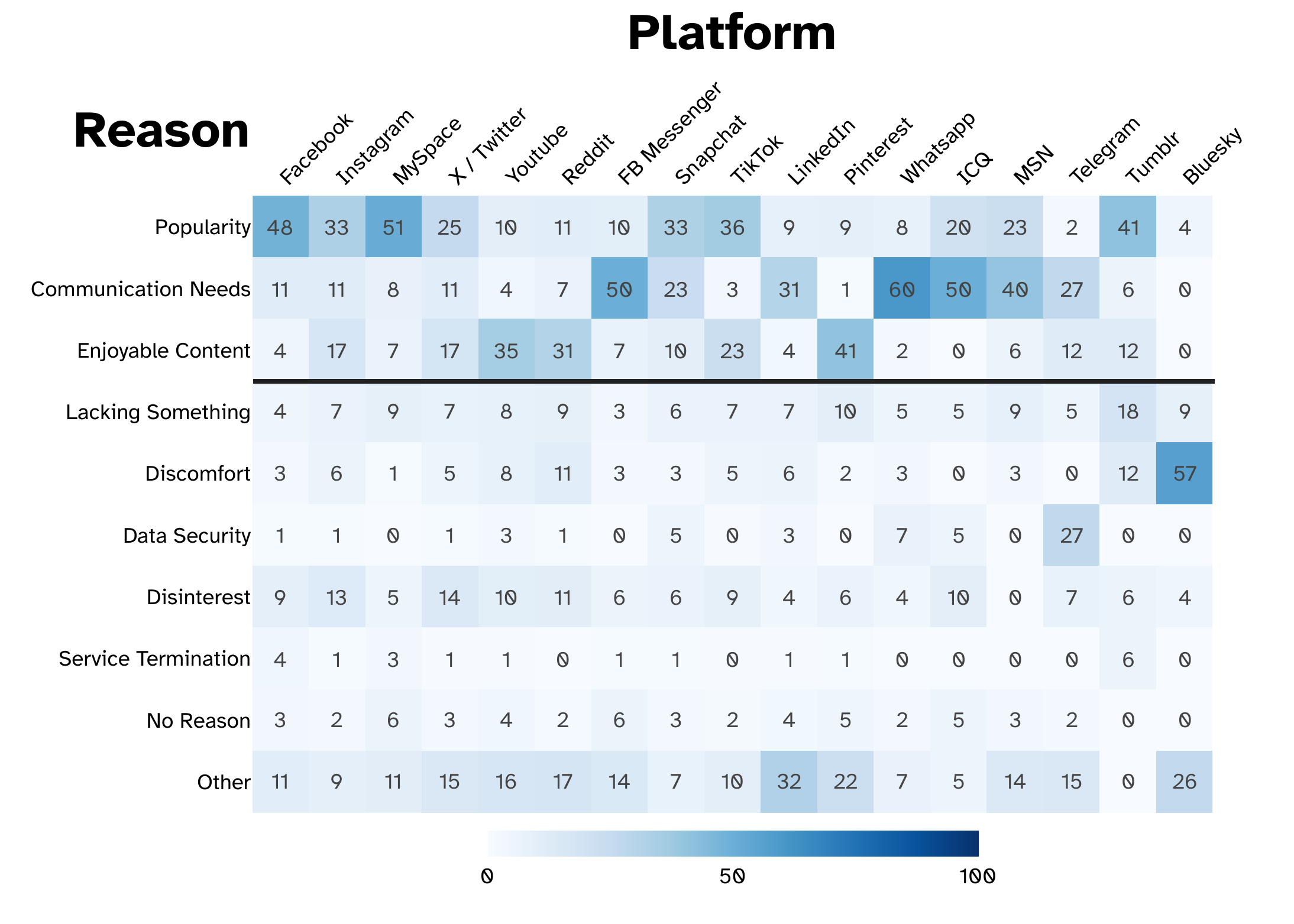}
    \caption{Pull Factor Distribution of each Platform: While most platforms have an overarching pull factor (top), external push factors
contribute to said pull (bottom)}
    \label{fig:heatmap-pull}
    \Description{Heatmap showing pull factor distributions across platforms. Platforms are listed along the top, and journey reasons along the left. Darker blue cells indicate higher pull rates. Popularity is the strongest pull factor overall, especially for Facebook, Instagram, and TikTok. Communication is dominant for platforms like ICQ, WhatsApp, and Messenger. Enjoyable content stands out on YouTube, Reddit, and Instagram. Smaller contributions come from reasons such as a lack of something, discomfort, data security, or disinterest. Service termination appears only in a few cases. An “Other” category is present across platforms, with Bluesky showing the highest relative rate there. The figure highlights that most platforms are primarily associated with one central pull factor, but additional factors also shape adoption.}
  \end{subfigure}
  \caption{Journey overview of all participants.}
  \label{fig:overall}
\end{figure*}

Each Journey is influenced by two factors: \textit{pull factors} that attract people toward a platform, and \textit{push factors} that drive them away from another. Figure~\ref{fig:heatmap-pull} shows this dynamic by distributing the number of Journeys towards a platform by different reasons. At the top of the figure are pull factors of the platform. At the bottom are push factors from other platforms. While most Transition edges can be interpreted as a push away from one platform and a pull toward another, Journeys also have natural beginnings and ends. In graph terms, some platforms appear only as entry points (outdegree without indegree), while others appear only as exit points (indegree without outdegree), reflecting that not all Transitions are continuous sequences.

This becomes clear in the Journeys \texttt{MySpace} $\rightarrow$ \texttt{Facebook} and \texttt{Facebook} $\rightarrow$ \texttt{Instagram}. These were the most common Journeys in our data. MySpace was first adopted for its popularity: ``[it seemed like] everyone was moving to MySpace''~(P462). But over time, its design began to push people away. Participants described slow load times~(``20 minutes to load''~P185), chaotic customization~(``everyone customized their pages [and it] was difficult loading [them]''~P140), and awkward features, such as MySpace’s public ranking of ``Top 8 friends'' ~(``I didn’t like the ‘8 favorite friends’''~P834). As one participant put it: ``MySpace became outdated''~(P580).

These pushes amplified the pull of Facebook. Journeys toward Facebook were motivated by social presence, as ‘more family and friends [were] over on Facebook’ (P708). Others stressed that they were able ``[to] connect with past friends''~(P629). Invitations reinforced this pull: ``I was invited by my daughter''~(P780), ``My mom got me hooked on the app''~(P550). For college students, it provided a dedicated hub: ``Facebook was initially for college students only, so it gave [them] a way to connect''~(P343). Facebook was popular and benefited from frustrations with MySpace.

Instagram repeated this pattern. Its pull came not only from popularity but also from its focus and simplicity. As P942 explained, ``My friends were more interested in uploading pictures than traditional posts.'' Participants contrasted this with Facebook, which P298 described as ``too many posts I didn’t want to see'' and as a place where ``I didn’t control the algorithm.'' Others echoed this sentiment, with P335 calling it ``a toxic political environment.'' By contrast, Instagram’s cleaner and more visual format promised relief. As P430 put it, ``[It] offered simpler, cleaner user experience while [Facebook] was becoming too cluttered``. 

At the same time, Instagram benefited from pushes from other platforms. Transitions mentioned larger shifts such as Yahoo’s acquisition of Tumblr~(P714) or Musk’s takeover of Twitter~(P610, P127, P93). For Twitter, every fourth Transition away from it was motivated by discomfort, half of which were related to the change of leadership. These events reinforced the pull of users who were dissatisfied elsewhere. Much like Facebook thrived on the decline of MySpace, Instagram consolidated its position by drawing strength from external pushes.  

Together, these Transitions show that adoption is not only about features or entertainment. It is also about how platforms fail their users and how those failures create openings for others. 

\subsection{How have platform design features and social factors influenced different generations throughout their Journeys? (RQ3)}
\label{sec:rq3}

Figure~\ref{fig:crossgen} visualizes social media Transitions across generations, with platforms categorized by their shape and source after Zhang et al., as detailed in Section~\ref{sec:methods-analysis}~\cite{zhangFormFrom2024}. As we use the framework to produce generalizable knowledge, a flow visualization captures the general Journey of each generation. We counted the number of Transitions between each Form-From category and mapped the counts to the arrow thickness. For each category, we annotated the platform mentioned the most for the respective Journey. We applied a cut-off of 5\% to highlight the most frequently occurring Journeys. Transitions from Flat Networks were predominantly from Facebook. About one in five of these Transitions (22\%) were toward Threaded Networks, predominantly to Instagram. The visualization helps us understand the overall Transitions for all generations and identifies broader themes on why people have used various social media platforms holistically. For every generation, we provide further nuance on push and pull dynamics with a heatmap for reason counts.

\subsubsection{Boomers: Familiarity, Family, and Withdrawal from Hostility}
\label{sec:boomers}

\begin{figure*}[t]
  \centering
  \begin{subfigure}[t]{0.49\textwidth}
    \includegraphics[width=\linewidth]{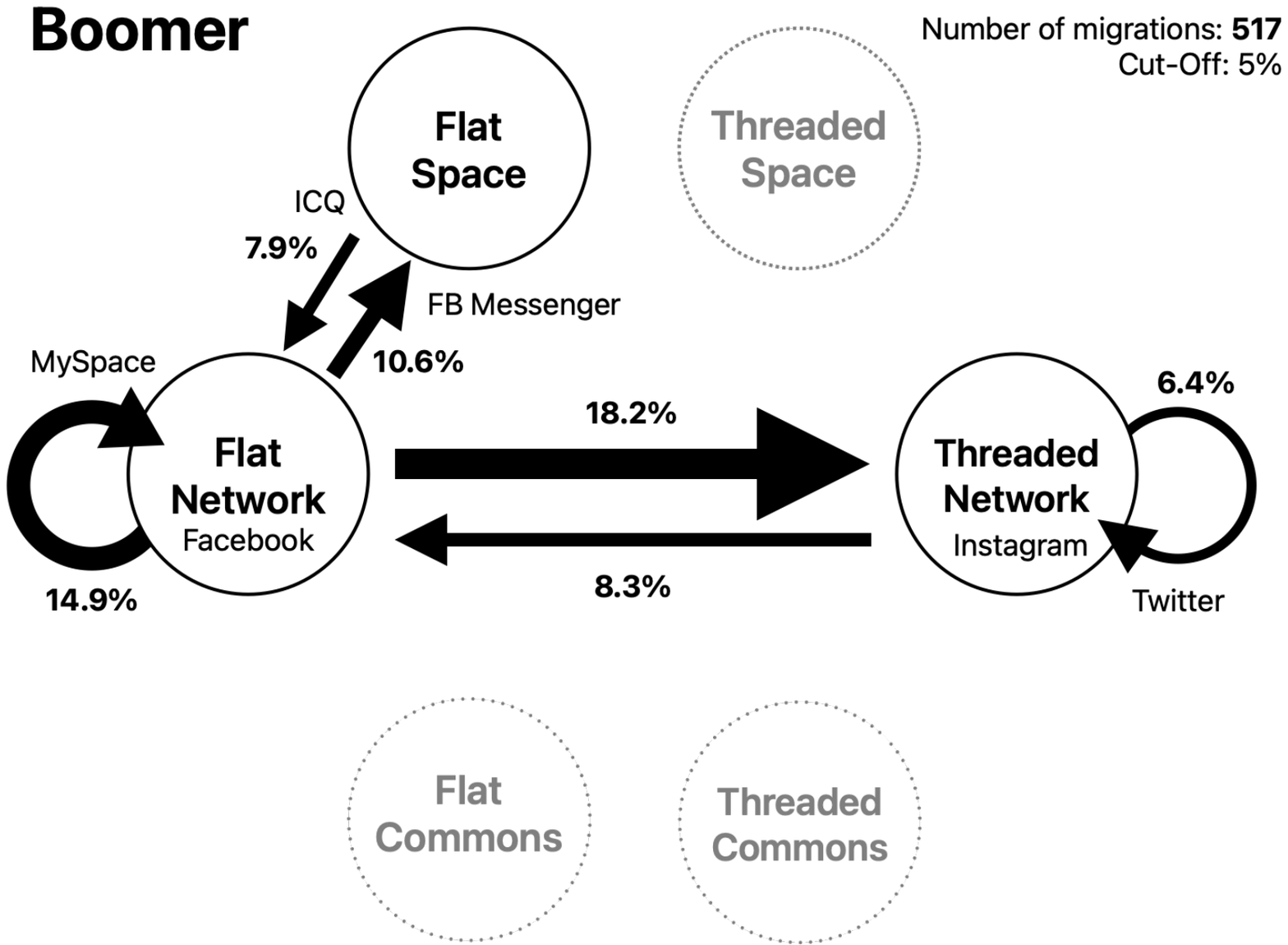}
    \caption{Distribution of Boomer Journeys across the Form--From design space. Most Journeys stayed within Flat Networks, highlighting the importance of familiarity and simplicity. Movements into Threaded Networks were less frequent and often motivated by withdrawal from stress elsewhere.}
    \label{fig:boomer-formfrom}
    \Description{The figure shows Transition flows specific to Boomer participants, with 517 Transitions included at a 5\% cut-off. Circles represent categories: Flat Network (Facebook, MySpace), Flat Space (Messenger, ICQ), and Threaded Network (Instagram, Twitter). Arrows indicate Transition directions with percentages. The largest flow is from Facebook to Instagram at 18.2\%, followed by a return flow of 8.3\%. MySpace has a self-loop of 14.9\%, while Facebook retains 10.6\%. Additional flows include 7.9\% from Facebook to ICQ and 6.4\% from Instagram to Twitter. The distribution highlights strong retention and movement within Flat Networks, with comparatively smaller shifts into Threaded Networks.}
  \end{subfigure}
  \hfill
  \begin{subfigure}[t]{0.49\textwidth}
    \includegraphics[width=\linewidth]{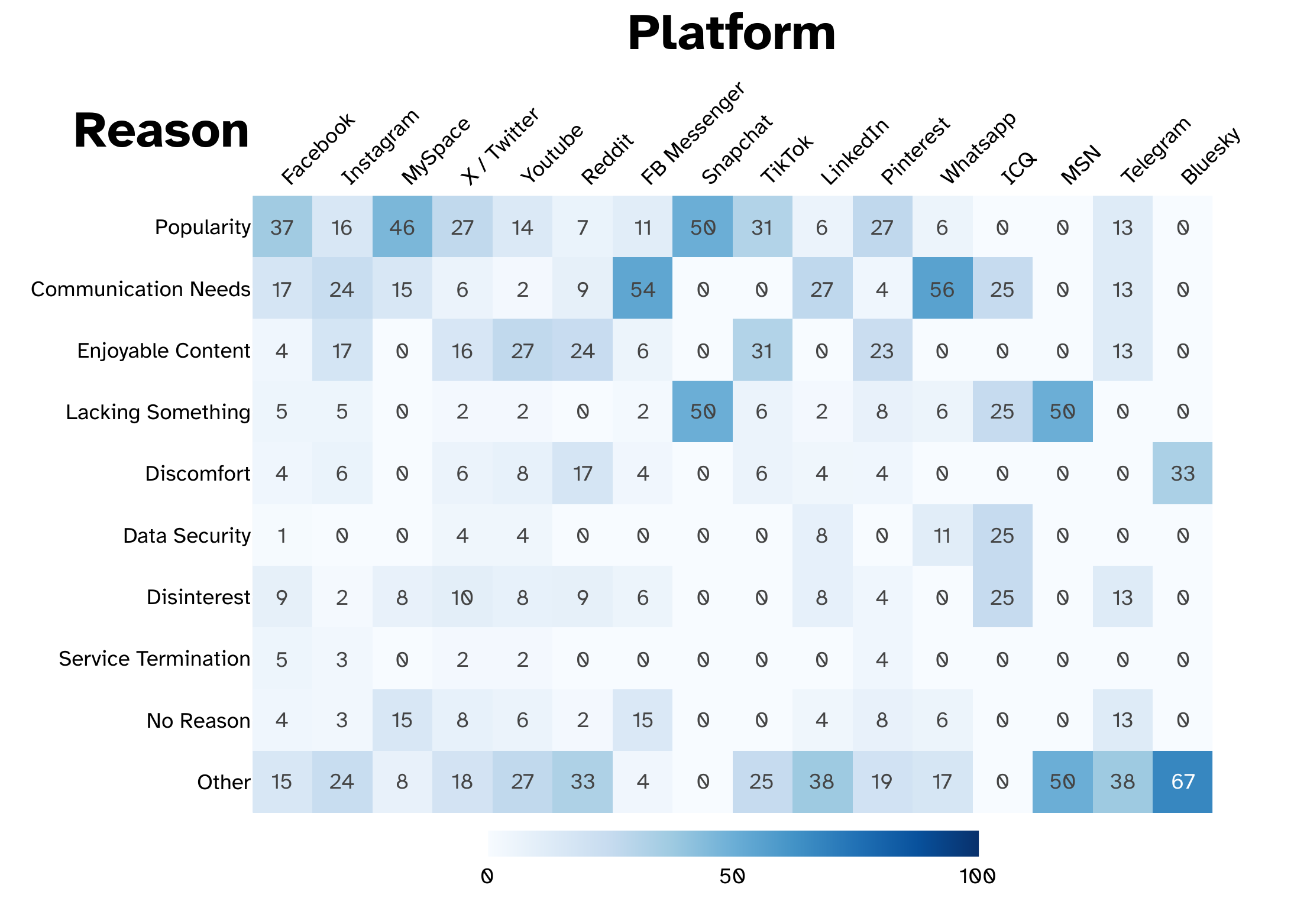}
    \caption{Push and pull factors for Boomer participants. Popularity and connection were the strongest pulls toward Flat Networks, while discomfort, disinterest, and political hostility pushed people away. Threaded Networks were often adopted as refuges from these pushes.}
    \label{fig:heatmap-boomer}
  \end{subfigure}
  \caption{Journey overview of the Boomer generation.}
  \label{fig:boomer}
  \Description{Heatmap of pull factors for Boomer participants across different platforms. Platforms are arranged along the top, and journey reasons along the side. Darker blue shading indicates higher pull rates. Popularity is the strongest factor overall, particularly for Facebook and Instagram. Communication also plays a significant role, especially for platforms like Messenger and ICQ. Enjoyable content shows moderate influence for YouTube and Reddit. Smaller contributions appear under discomfort, disinterest, and data security, with a noticeable value for Bluesky under discomfort. The “Other” category is spread across multiple platforms, with Bluesky showing the highest value. The visualization highlights that Boomers were most strongly drawn to platforms for popularity and connection, while other reasons were secondary.}
\end{figure*}

Boomers~(203 participants, born 1946--1964) account for 517 Journeys~(Figure~\ref{fig:boomer}). Their Journeys were shaped by three themes: \textit{familiarity}, \textit{family}, and \textit{withdrawal}. 
These themes summarise how Boomer participants moved across platforms: turning toward services that felt familiar, joining spaces where close ties were already present, and leaving platforms when political or emotional strain became too strong.

Flat Networks (i.e., platforms with linear feeds shared among contacts) promised a sense of familiarity. Meanwhile, Threaded Networks (i.e., platforms with threaded feeds shared among contacts) were adopted more reluctantly, often as a response to political or emotional stress. Professional platforms were abandoned when they no longer aligned with a life stage. Across the sample, Boomer participants showed a preference for stability and familiarity, with movements shaped by existing relationships and by avoidance of stressful or hostile environments.

Adoption of Facebook illustrates how Boomers entered social media through familiar design and trusted ties. The heatmap shows that popularity was a strong pull factor. However, participants also highlighted connection. P785 described joining Facebook because of ``old friends and family,'' while P780 noted: ``I was invited by my daughter''. Others mentioned the promise of ``more and better ways to reconnect with long lost friends''~(P568). These Journeys show how family and connection drove adoption.  

At the same time, Facebook became a site of withdrawal. Push factors such as discomfort and disinterest appear clearly in the heatmap. For Boomers, this was tied to the rise of politics and algorithmic feeds. P335 describes that ``Facebook has become a toxic political environment,'' while P435 complained how ``Facebook is manipulating its users' feeds for their own purposes, and not to the benefit of them or society in general''. Here, design features such as algorithmic curation transformed what had been a site of belonging into one of alienation.  

Adoption of Threaded Networks, such as Instagram and Twitter, was less about attraction and more about seeking refuge. Journeys into these platforms were often framed as an escape from stress elsewhere. P693, who moved after leaving MSN, explained: ``I wanted to get away from headline news for the most part. [Political] story lines were becoming too stressful''. Threaded Networks thus became a place to step away from negativity.  

Some Journeys reflected life transitions. Leaving LinkedIn, for example, was connected to retirement. P800 described: ``I retired and wanted to move away from people who were still working, as I wanted to address other issues in my life''. These movements show how changes in life stage shaped platform use, with participants stepping away from professional spaces once they no longer matched their daily lives.

\subsubsection{Gen X: Trust and Interests}

\begin{figure*}[t]
  \centering
  \begin{subfigure}[t]{0.49\textwidth}
    \includegraphics[width=\linewidth]{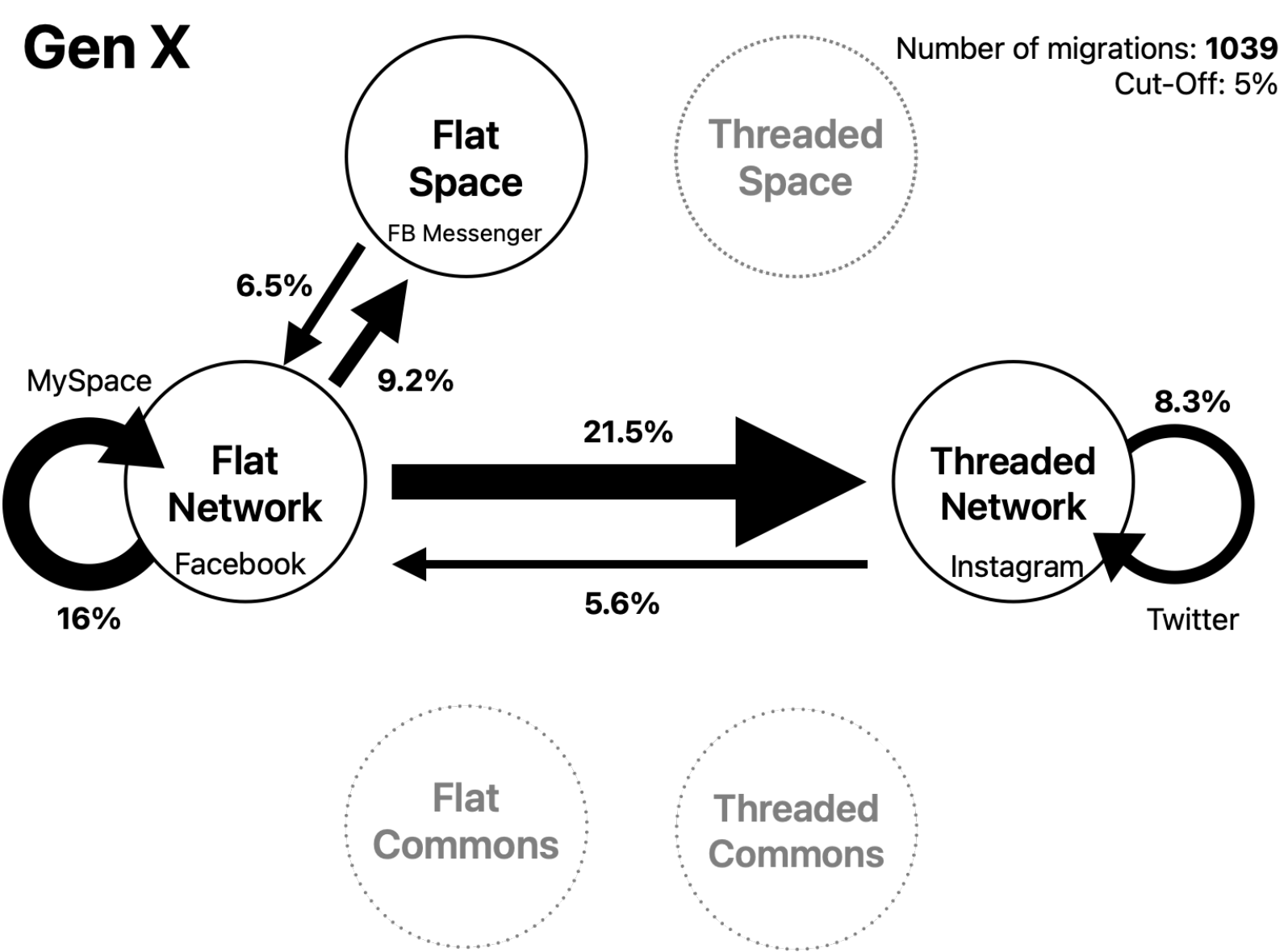}
    \caption{Distribution of Gen~X Journeys across the Form--From design space. Gen~X spread more evenly across Flat and Threaded Networks. While Facebook anchored many Journeys, ad fatigue and privacy concerns led to frequent adoption of alternatives such as Instagram, Reddit, and Twitter.}
    \label{fig:genx-formfrom}
    \Description{Transition flows for Gen X participants, with 1,039 Transitions included at a 5\% cut-off. Circles represent platform categories: Flat Network (Facebook, MySpace), Flat Space (Messenger), and Threaded Network (Instagram, Twitter). Facebook is central, with a 21.5\% flow to Instagram and a 5.6\% return flow. MySpace shows a 16\% self-loop, while Facebook retains 9.2\%. Instagram retains 8.3\%. Smaller flows include 6.5\% from Facebook to Messenger. The distribution indicates Gen X moved more evenly across Flat and Threaded Networks, with Facebook anchoring many journeys but notable Transitions toward alternative platforms.}
  \end{subfigure}
  \hfill
  \begin{subfigure}[t]{0.49\textwidth}
    \includegraphics[width=\linewidth]{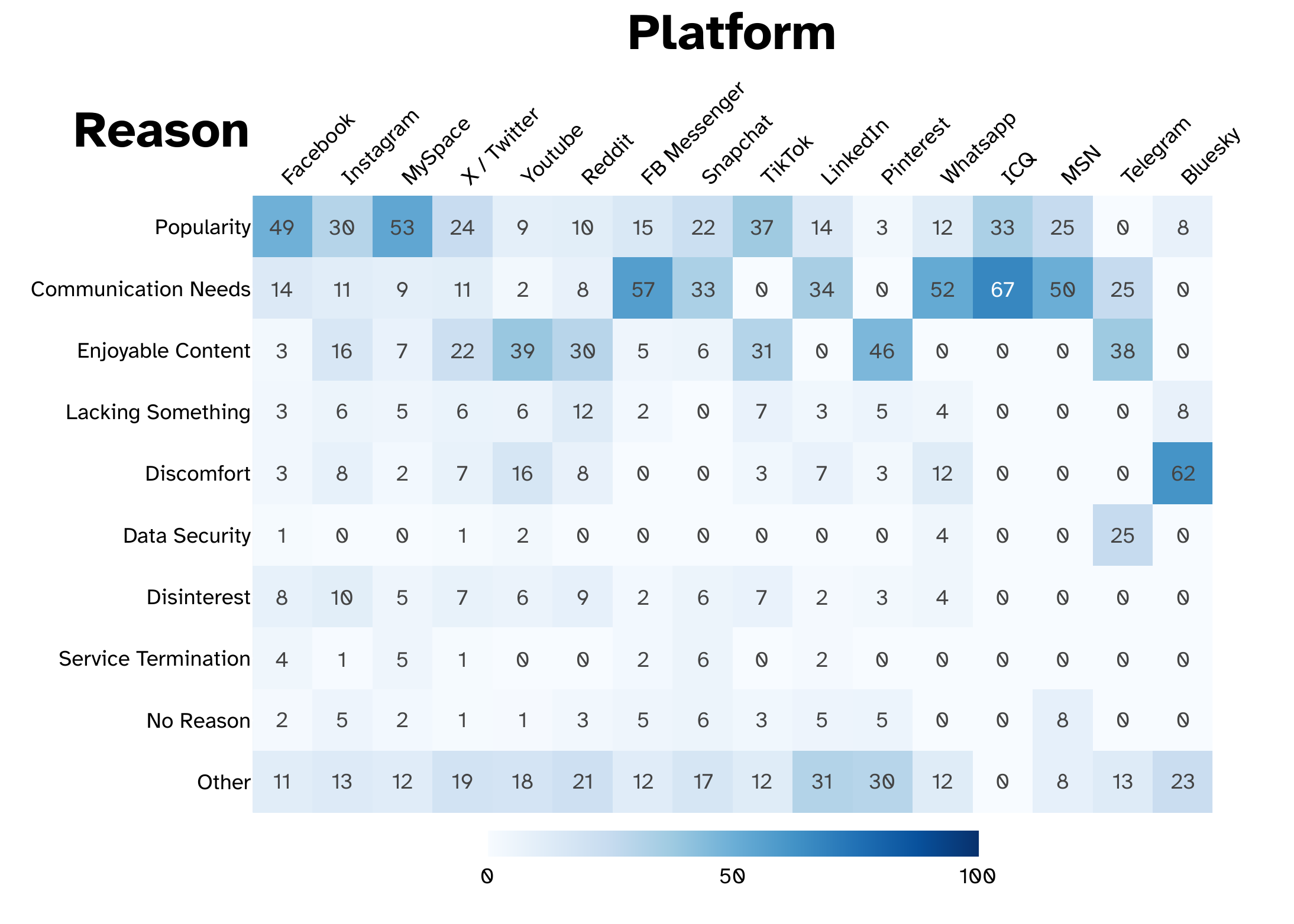}
    \caption{Push and pull factors for Gen~X participants. Pull factors such as popularity, community, and richer content drove the adoption of new platforms. Push factors such as censorship concerns, privacy issues, and frustration with ads pushed people away from older ones.}
    \label{fig:heatmap-genx}
  \end{subfigure}
  \caption{Journey overview of the Gen~X generation.}
  \label{fig:genx}
  \Description{Heatmap of pull factors for Gen X participants across platforms. Platforms are listed along the top, and journey reasons along the side. Darker blue shading represents higher pull rates. Popularity is the strongest factor overall, especially for Facebook, Instagram, and TikTok. Enjoyable content plays a major role for YouTube, Reddit, and Instagram. Communication is important for Messenger and ICQ. Other reasons, such as discomfort, disinterest, and data security, appear less frequently, though discomfort is notable for Bluesky. An “Other” category is present across platforms, with higher values for Pinterest, ICQ, and Bluesky. Overall, Gen X participants were mainly drawn by popularity, community, and richer content, while additional factors shaped platform-specific adoption.}
\end{figure*}

Gen X~(303 participants, born 1965--1980) account for 1,039 Journeys~(Figure~\ref{fig:genx}). Their Journeys followed two themes: \textit{trust} and \textit{interests}. 
These themes summarise how Gen X participants moved across platforms: reacting to concerns about censorship, leadership, and privacy, while also seeking out spaces that matched their hobbies, preferred content, or informational needs.

Distrust in leadership, censorship, and privacy concerns pushed Gen X participants away from older platforms, while curiosity, hobbies, and entertainment pulled them toward new ones. Across the sample, Gen X participants balanced scepticism with exploration, using social media both as a place to express concerns and as a resource for deeper content and learning. They used platforms as outlets for expression as well as resources for deeper content and learning on sites like YouTube, Pinterest, and Reddit.

Adoption of Facebook was initially driven by popularity and familiarity, but trust quickly became an issue. Participants described leaving after feeling manipulated or silenced. P975 explained: ``I left when censorship started becoming a problem''. Others framed their adoption of Twitter in a similar fashion. P869 stated: ``I needed a place to voice my distress about politics,'' while P988 looked for ``less censorship and more free speech''. These Journeys show how adoption was tied to distrust in leadership and algorithmic curation, with Threaded Networks serving as outlets for expression. In terms of privacy, P854 described leaving Facebook for Reddit because they had ``lost trust in the leadership and felt [their] privacy was not secure''.

At the same time, Gen X also adopted platforms around content depth and personal interests. Prevalent Journeys toward YouTube, Pinterest, and Reddit highlight this. P934 noted how TikTok gave them DIY repair videos. P932 moved from Pinterest to YouTube because ``YouTube had better, more in-depth content''. Similarly, P260 explained moving from Pinterest to Reddit for ``deeper content in specific subjects and people with similar interests''. Together, these movements show how Gen X used platform features and content depth to find spaces that aligned with their interests while stepping away from platforms they found untrustworthy or intrusive.

\subsubsection{Millennials: Platform Specialization and Identity Management}
\label{sec:millennials}

\begin{figure*}[t]
  \centering
  \begin{subfigure}[t]{0.49\textwidth}
    \includegraphics[width=\linewidth]{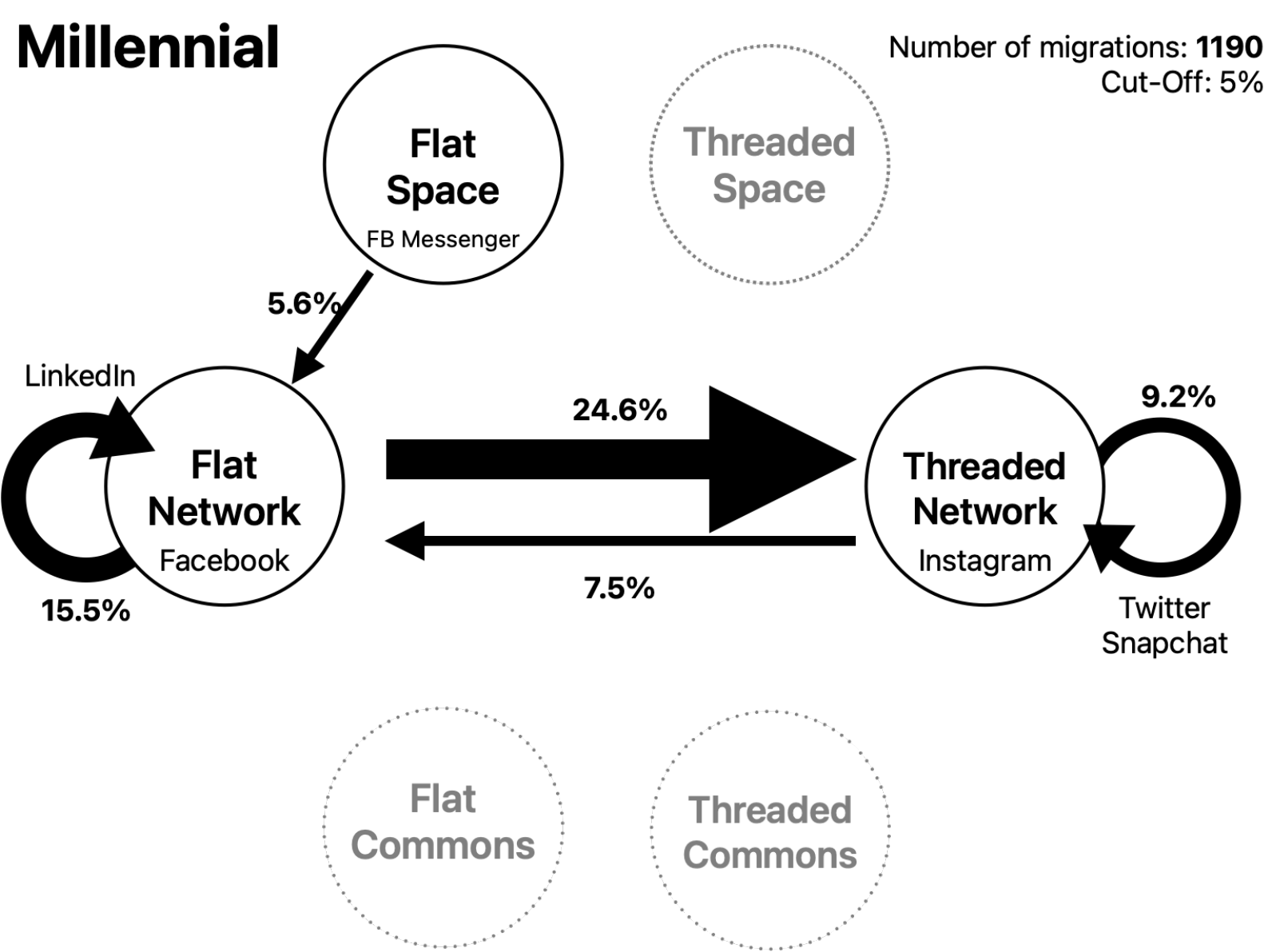}
    \caption{Distribution of Millennial Journeys across the Form--From design space. Facebook anchored early use, but frequent moves toward Messenger, LinkedIn, and Instagram show how participants divided social, professional, and personal interactions.}
    \label{fig:millennial-formfrom}
    \Description{The figure shows Transition flows for Millennial participants, based on 1,190 Transitions with a 5\% cut-off. Circles represent categories: Flat Network (Facebook, LinkedIn), Flat Space (Messenger), and Threaded Network (Instagram, Twitter, Snapchat). The largest flow is from Facebook to Instagram at 24.6\%, with a 7.5\% return from Instagram to Facebook. Facebook retains 15.5\% of users, while Instagram retains 9.2\%. Additional flows include 5.6\% from Facebook to Messenger. The distribution highlights Facebook’s role as an early anchor, with Millennials frequently moving to Instagram for personal interaction and to LinkedIn and Messenger for professional or social purposes.}
  \end{subfigure}
  \hfill
  \begin{subfigure}[t]{0.49\textwidth}
    \includegraphics[width=\linewidth]{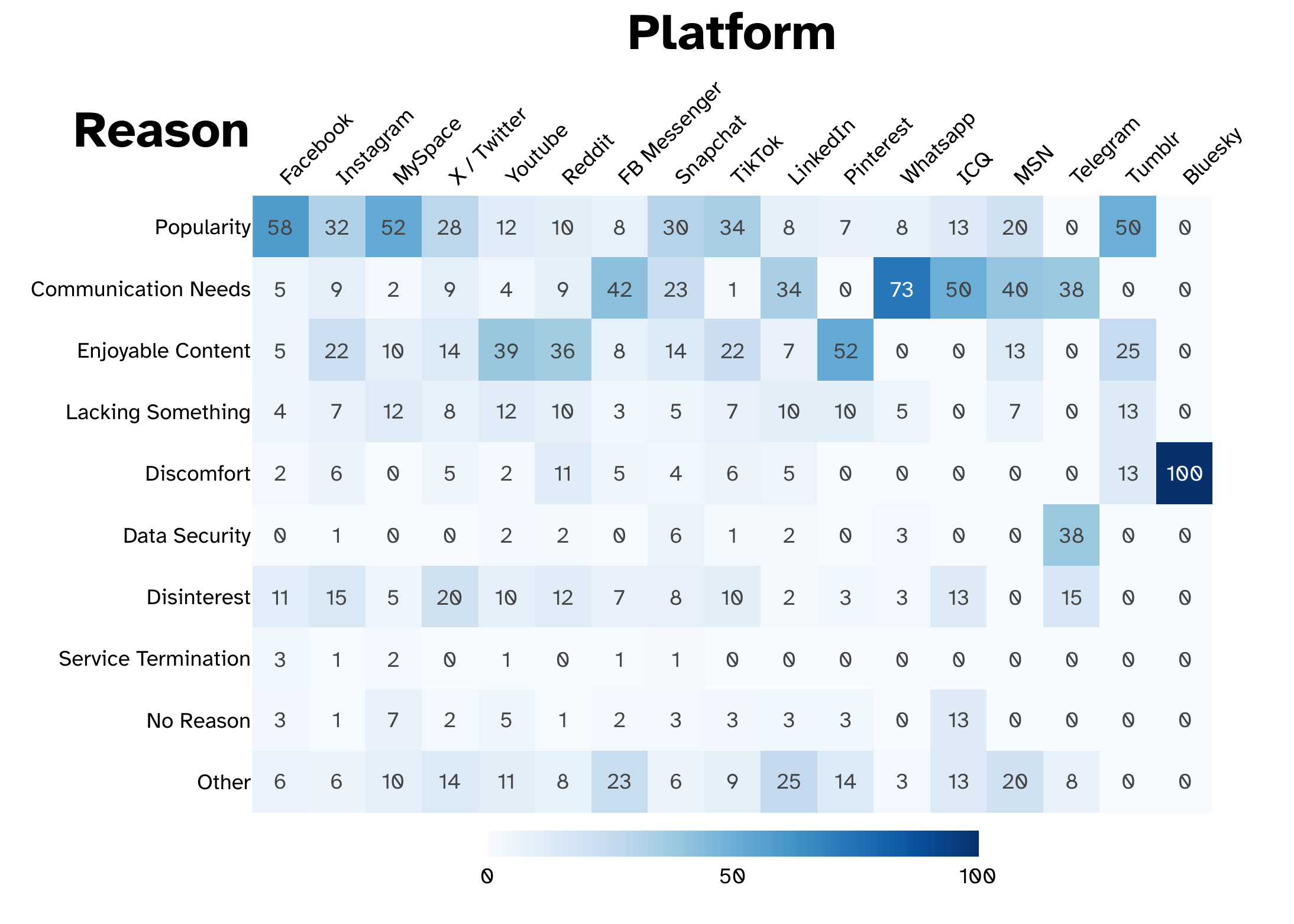}
    \caption{Push and pull factors for Millennial participants. Popularity anchored Facebook, but disinterest~(11\%) and intergenerational crowding led to shifts. Career-building~(15.5\% toward LinkedIn) and entertainment~(9.2\% within Instagram/Twitter/Snapchat) structured further adoption.}
    \label{fig:heatmap-millennial}
  \end{subfigure}
  \caption{Journey overview of the Millennial generation.}
  \label{fig:millennial}
  \Description{The figure shows a heatmap of pull factors for Millennial participants across different platforms. Platforms are listed along the top, and journey reasons along the side. Darker blue indicates stronger pull rates. Popularity dominates overall, especially for Facebook, Instagram, and TikTok. Enjoyable content is important for YouTube, Reddit, and Instagram, while communication is a strong factor for Messenger and WhatsApp. Disinterest is notable for Facebook, reflecting 11\% of shifts. Career-oriented moves are visible in LinkedIn, linked to 15.5\% of activity. Entertainment-driven adoption appears in Instagram, Twitter, and Snapchat, showing 9.2\% combined. Other reasons are distributed across platforms, with moderate pull toward Pinterest, ICQ, and Bluesky. The visualization highlights how Millennials balanced popularity with career-building and entertainment as primary drivers.}
\end{figure*}

Millennials~(298 participants, born 1981--1996) account for 1,190 Transitions~(Figure~\ref{fig:millennial}). Their Journeys followed three themes: \textit{platform specialization}, \textit{identity management}, and \textit{entertainment}. 
These themes summarise how Millennial participants moved across platforms: assigning different roles to different platforms, managing audiences across personal and professional contexts, and turning to visual or lightweight services for entertainment.

Millennials use social media in a specialized and purposeful way. Across the sample, Millennial participants divided their online lives across platforms rather than relying on a single service, tailoring each to specific social, professional, or leisure needs. They used Facebook as an anchor, Messenger for private communication, LinkedIn for professional identity, and Instagram, Twitter, or Snapchat for entertainment. Their adoption reflects careful audience and identity management, with entertainment platforms offering lightweight content. Taken together, Millennials navigated social media by tailoring each platform to a specific social, professional, or leisure need.

Facebook was the dominant entry point~(58\% citing popularity). Yet Transitions away were frequent: 24.6\% of Journeys went to Threaded Networks like Instagram, and 5.6\% toward Flat Spaces like Messenger. Participants explained this Journey as a way to narrow audiences: ``Facebook was becoming too big and older [family members] were now on the platform. I did not want to share certain things with my grandma''~(P044). 

Identity management was visible in professional adoption. More than one in seven Journeys connected Facebook to LinkedIn, marking a career entry. P746 explained: ``I wanted to build a professional page,'' while P545 described LinkedIn as necessary once they ``started [their] career in earnest''. LinkedIn thus layered a professional identity on top of earlier personal and social use.  

Entertainment played an important role as well. Instagram and Snapchat were adopted for filtered, visual-first, lightweight content, with one in 10 Journeys encompassing Instagram, Twitter, and Snapchat. Some participants were drawn by celebrity interaction on Twitter~(P257), while others noted Instagram’s more enjoyable media. Across these platforms, entertainment served as a flexible space separate from family or professional audiences, reinforcing the pattern of dividing roles across services.

\subsubsection{Gen Z: Control and Well-Being}
\label{sec:genz}

\begin{figure*}[t]
  \centering
  \begin{subfigure}[t]{0.49\textwidth}
    \includegraphics[width=\linewidth]{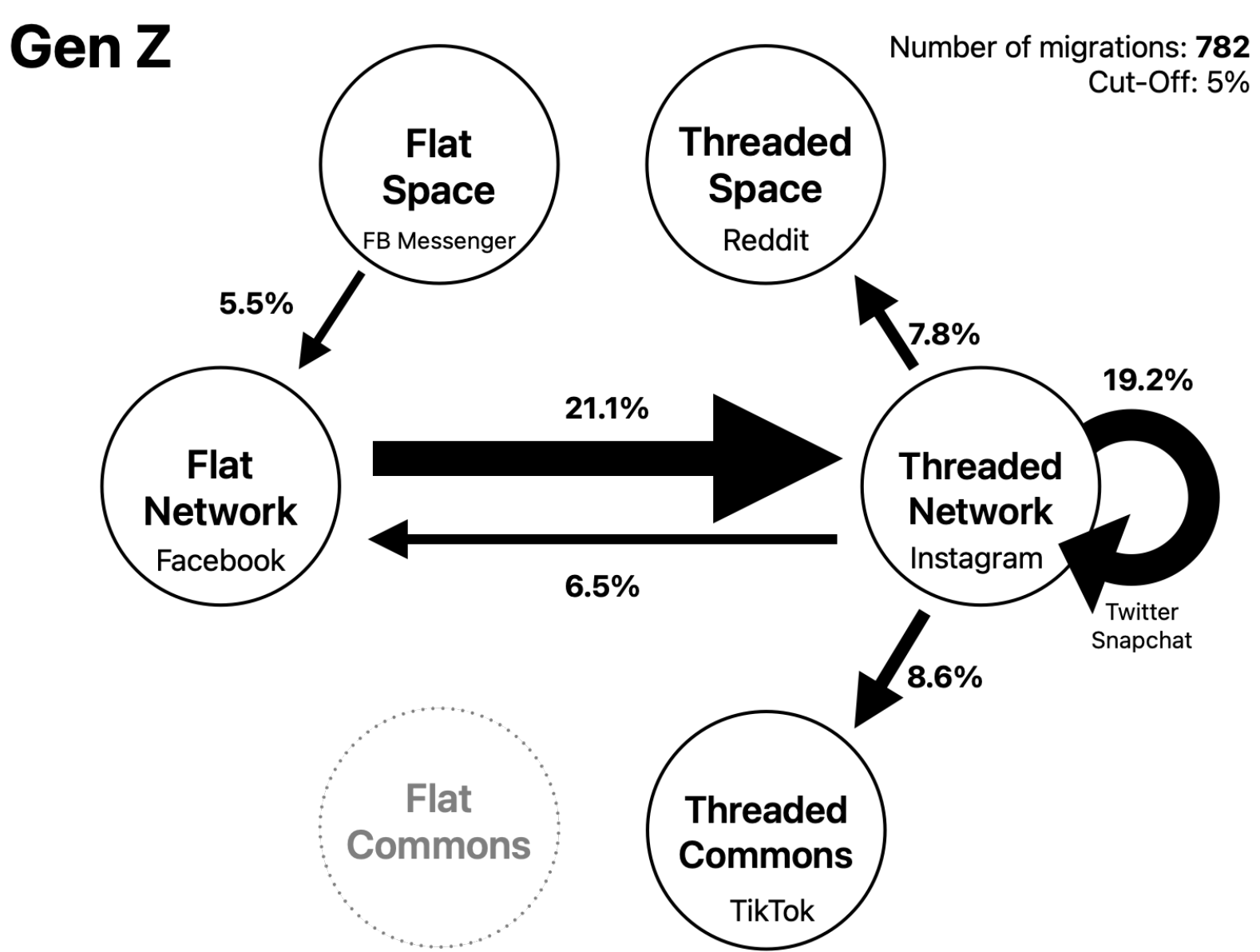}
    \caption{Distribution of Gen~Z Journeys across the Form--From design space. Journeys were fragmented across Instagram~(19.2\%), TikTok~(8.6\%), Reddit~(7.8\%), and YouTube, with no single dominant hub.}
    \label{fig:genz-formfrom}
    \Description{Transition flows for Gen Z participants, based on 782 Transitions with a 5\% cut-off. Circles represent categories: Flat Network (Facebook), Flat Space (Messenger), Threaded Network (Instagram, Twitter, Snapchat), Threaded Space (Reddit), and Threaded Commons (TikTok). Instagram is the largest hub, retaining 19.2\% and receiving 21.1\% from Facebook, while 6.5\% move back to Facebook. TikTok accounts for 8.6\% of flows, Reddit for 7.8\%, and Messenger for 5.5\%. The distribution highlights a fragmented pattern, with no single dominant platform, as Gen Z spreads activity across multiple spaces.}
  \end{subfigure}
  \hfill
  \begin{subfigure}[t]{0.49\textwidth}
    \includegraphics[width=\linewidth]{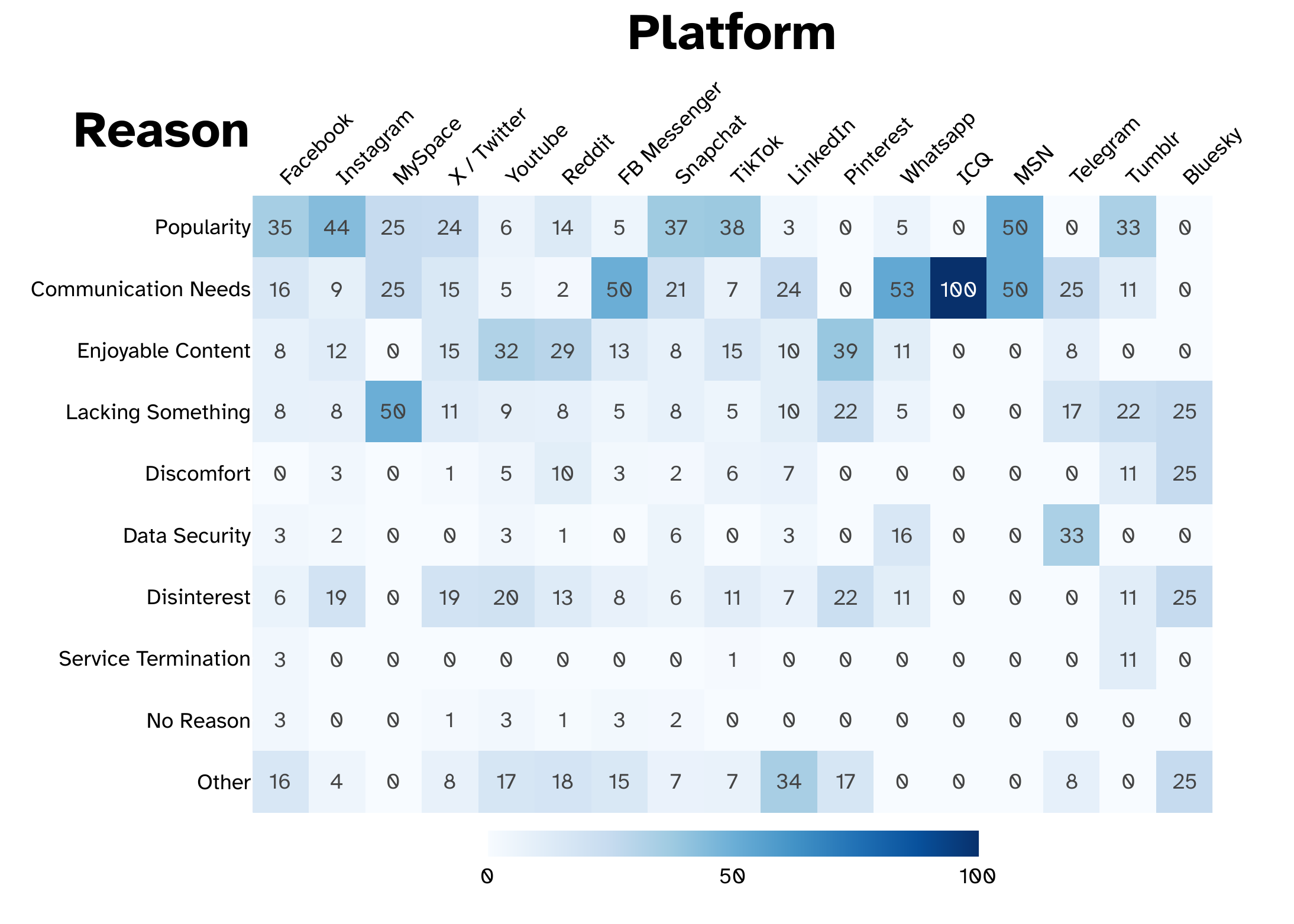}
    \caption{Push and pull factors for Gen~Z participants. Popularity played a role~(e.g., Instagram 44\%), but disinterest~(19\%) and emotional discomfort~(e.g., Instagram comparisons, Twitter anxiety) drove many shifts.}
    \label{fig:heatmap-genz}
  \end{subfigure}
  \caption{Journey overview of the Gen~Z generation.}
  \label{fig:genz}
  \Description{Heatmap of pull factors for Gen Z participants across platforms. Platforms are listed along the top, and journey reasons along the side. Darker blue shading represents stronger pull rates. Popularity is the leading factor, especially for Instagram (44\%). Enjoyable content is also influential, with higher values for YouTube, TikTok, and Reddit. Disinterest stands out at 19\%, particularly linked to Facebook and Twitter. Discomfort is another prominent factor, with notable values for Instagram and Twitter, reflecting emotional strain. Other reasons are present across multiple platforms, with moderate contributions toward TikTok and Bluesky. The visualization highlights that while popularity remains important, Gen Z shifts are also strongly shaped by disinterest and discomfort.}
\end{figure*}
Gen Z~(196 participants, born 1997--2012) account for 782 Transitions~(Figure~\ref{fig:genz}). Their Journeys followed two themes: \textit{control} and \textit{well-being}. 
These themes summarise how Gen Z participants moved across platforms: curating what they see, resisting algorithmic pressure, and avoiding spaces that felt stressful or harmful.

Gen~Z moved across platforms to manage exposure, preserve emotional well-being, and access valued content. Across the sample, Gen Z participants made active decisions to shape what they encountered online, foregrounding control over feeds and attention as part of safeguarding their well-being. While TikTok offered novel, short-form entertainment, it played an ambivalent role, leading some to move to alternatives like YouTube or Reddit for more control, niche content, or safer experiences. Overall, Gen Z adoption reflects careful management of exposure, content, and personal well-being.

Control often meant curating content and resisting algorithmic influences. P640 explained: ``I wanted more control over the content I was seeing rather than being primarily fed it by an algorithm''. In their Journeys, 21.1\% moved from Facebook to Instagram, while others turned toward TikTok~(8.6\%) and Reddit~(7.8\%) for specialized feeds. A smaller number of participants used platforms functionally, e.g., accessing marketplaces~(P966, P914).  

Well-being concerns were even more pronounced. Every fifth Journey to Instagram and Twitter was driven by disinterest. However, participants also voiced discomfort on these platforms. P875 described Instagram as making them ``compare [themselves] to other people,'' while P022 left Twitter because it ``made [them] anxious''. Exposure to toxic content also played a role with ``hate speech targeted towards minority groups'',~(P965). These negative experiences often motivated a Transition toward platforms perceived as safer or more positive.  

Threaded Commons (i.e., a platform where content is shared platform-wide with threaded conversations), exclusively being TikTok, was most frequently reported by Gen Z participants. Participants described its novelty as unmatched: ``At the time, TikTok was the only one to have the short form, scrollable content, and it was unmatched''~(P042). Others used it briefly but abandoned it for being ``too political''~(P918), ``very annoying''~(P042), or ``a waste of time''~(P678). Some contrasted it directly with alternatives, e.g., moving to YouTube because it had ``better longer video content''~(P154) or to Reddit because they ``preferred reading over watching''~(P655). Still others continued using it in tandem for light entertainment~(P022). These Transitions position TikTok as both highly frequented and divisive among Gen Z Journeys. Participants split over whether TikTok’s fast-paced, platform-wide content was uniquely engaging or simply overwhelming.

Gen~Z Journeys leveraged Reddit as a way to combat algorithmic curation. As P640 explained: ``I wanted more control over the content I was seeing rather than being primarily fed it by an algorithm''. Others emphasized its niche and informational value, from ``communities of a specific niche like couponing''~(P721) to ``answers to questions [they are] afraid to ask''~(P550). For some, Reddit’s appeal lies in its breadth of authentic perspectives: ``Reddit has so many communities for literally everybody, and it's much easier to find a community and read what others have posted''~(P040).
Together, these patterns show how Gen Z relied on platform changes to fine-tune their online environments, prioritising control and well-being throughout their Journeys.

\subsubsection{Cross-Generational Patterns}
\label{sec:crossgen}
Among the 3,553 collected Journeys, the most prominent one was \texttt{Flat Network} $\rightarrow$ \texttt{Threaded Network} with Facebook$\rightarrow$Instagram as its main representative. Participants reported using both platforms at the same time~(P487, P700). In Journeys within the same design space, such as Flat Network$\rightarrow$Flat Network, the most frequent one was \texttt{MySpace} $\rightarrow$ \texttt{Facebook}. \texttt{Threaded Network} $\rightarrow$ \texttt{Threaded Network} Journeys included various Journeys between Instagram, Twitter, YouTube, and Snapchat. The wide majority of Journeys toward Threaded Spaces were directed at Reddit. Journeys into Flat and Threaded Commons were fewer and fell under the cut-off.  

\section{Discussion}

Our analysis shows that people rarely adopt or abandon platforms in isolation but rather in relation to the promises and shortcomings of what they already use. This corroborates and extends Fiesler and Dym’s study of fandom migrations, where users evaluated new platforms like AO3 or Dreamwidth on features and their ability to preserve community practices carried over from Tumblr~\cite{fieslerMoving2020}. In our data, similar relational judgments appear: MySpace’s clutter and slowness pushed users toward Facebook, while Instagram’s simplicity and visual focus offered relief from Facebook’s clutter. Social Media Journeys thus highlight that adoption is a comparative and community-shaped process.

Jungselius and Weilenmann identified three broad shifts in social media use: from active production to passive consumption, from public performance to private interaction, and from fun to problematic use~\cite{jungselius2025long}. Our findings corroborate these trajectories from a new methodological perspective by revealing that they unfold differently across generations. While active production was not a significant factor in our data, passive consumption emerged as a dominant feature across all cohorts, underscoring its centrality in contemporary social media practices. The shift from public performance to private interaction also diverges generationally: Millennials remain tied in Facebook while actively managing communication by separating professional and social spheres, whereas Gen Z fragments participation across multiple platforms to protect well-being during consumption. Similarly, the move from fun to problematic use manifests differently: for Gen X, curiosity, hobbies, and entertainment drew them toward new platforms, while distrust in leadership, censorship, and privacy concerns pushed them away from older ones; for Gen Z, concerns centered on balancing well-being with the pursuit of enjoyable content. These findings situate the shifts described by Jungselius and Weilenmann within a generational frame, demonstrating that such shifts are not uniform trends but distinct, negotiated strategies of engagement. Users actively curate their social media experience, balancing entertainment with well-being.

This notion of curation resonates with Buss et al.’s work on transgender users, who tailor identity presentation, manage disclosure risks, and shape feeds to affirm their identities~\cite{buss2022transgender}. Across generations, participants in our study similarly adjusted their social media practices to protect well-being and manage visibility. Millennials compartmentalized professional and social spheres while maintaining a Facebook presence, Gen Z fragmented platform use to reduce stress, and Boomers and Gen X selectively withdrew to sidestep information overload. Unlike Buss et al.’s participants, for whom identity work and disclosure were central, our participants framed curation practices around broader concerns, including attention, privacy, and mental health. By integrating these perspectives, our findings show that strategies of moderation and selective participation extend beyond high-stakes identity negotiation: they are generationally distinct responses to social media landscape, societal pressures, and the ongoing negotiation between enjoyment and well-being. DeVito's Personal Social Media Ecosystem~\cite{deVitoTooGay2018}, Haimson's Transition Machinery~\cite{haimsonTransitionMachinery2018}, and Kender and Spiel's~\cite {kender2025machine} Marginalization Machine all describe specialized scenarios that coincide with the experiences of our quota-based sample.

Mapping these Journeys onto the Form–From framework~\cite{zhangFormFrom2024} revealed that the historical evolution of design spaces (e.g., \texttt{Flat Space} $\rightarrow$ \texttt{Threaded Space} $\rightarrow$ \texttt{Flat Network} $\rightarrow$ \texttt{Threaded Network} $\rightarrow$ \texttt{Threaded Commons}) is not mirrored by the general Social Media Journeys of our participants. Even for the Boomer generation, Flat and Threaded Networks were central and most-frequented while other design spaces were explored for their communication needs (Flat Spaces with FB Messenger), their content quality (Threaded Spaces with Reddit), or content presentation~(Threaded Commons with TikTok). It also has to be noted that, contrary to the Reuters Digital News Report of 2025~\cite{reuters2025}, our participants rarely mentioned TikTok, which was the sole representative for Threaded Commons. Even though the platform is popular among Gen Z participants, its overall presence is not as pronounced as platforms owned by Meta, which has already implemented modern features that would be classified as Threaded Commons. 

\subsection{Why do we use Social Media?}
Our results show that the reasons behind platform adoption are far more complex and intertwined. General judgments about ``social media'' risk oversimplification: just because one platform suffers from poor moderation or misuse does not justify condemning all platforms. While concerns over harmful content are legitimate, they cannot erase the emancipatory potential these platforms hold for connection, self-expression, and community-building. In light of platforms introducing new features in relation to generative artificial intelligence~\cite{Lu_2023}, new platforms emerging as alternatives to Twitter~\cite{quelle2025academicsleavingtwitterbluesky}, and policies being discussed regarding the ban of social media platforms in various contexts~\cite{Australia2025,UK_Online_Safety_Act_Age_Assurance_2025, Guardian_OnlineSafetyAct_FreeSpeech_2025}, we need powerful tools to understand powerful technologies. Social Media Journeys enable us to distill nuance in seemingly simple acts of moving from one platform to another, and to uncover the reasons that make Transitions meaningful.

\subsection{Design, Research and Policy Implications}

We present seven design implications to inform the implementation, research, and policing of social media platforms.

\textbf{User experience happens in relation to competitors:} 
Our findings show that people use social media in many different ways, and that what most individuals have experienced online is unique, echoing social media's nature of collapsing contexts~\cite{boyd2014itscomplicated}. They adopt and abandon platforms in comparison with what they already know, judging new services against the failures and affordances of their predecessors. Each adoption is situated within a larger complex of alternatives. This means that platform experience itself is relational, shaped as much by histories and comparisons as by immediate features. A timely feature may be convenient to a certain type of person, but it rarely captures the comparative, historical, and community-shaped nature of platform experience. Platform requirements need to be defined well in consideration of why a platform matters in the broader social media landscape.

\textbf{Socialization Happens Within Contexts:}  
A single platform rarely captures the full scope of someone’s social media life. The same person who shares hobbies and debates anonymously on Reddit may also carefully manage their professional persona on LinkedIn. These social media spaces are not interchangeable but interdependent: choices made in one context are informed by the norms, risks, and expectations of another. Social media use is therefore best understood as a negotiation across overlapping social and professional worlds. Documenting what someone does on one site may be meaningful for capturing localized practices, but it falls short when the goal is to build generalizable knowledge. Users’ accounts are always situated within a broader ecosystem of competing and overlapping platforms, each tied to different social, professional, and economic logics~\cite{marwick2011tweet}. To recognize the complexity of social media use, designers, researchers, and policymakers must therefore approach Journeys as contextualized and interdependent rather than discrete, and situate findings within the multiplicity of environments in which communication unfolds.

\textbf{Platform Adoption Has Generational Nuance:} 
Generational differences are not accidental but shaped by historical context: different generations entered the social media landscape at distinct points and carry forward unique expectations, habits, and strategies. Treating ``youth'' or ``users'' as homogeneous groups obscures these distinctions and risks misinterpreting behaviors. Our findings speak to the deeper meaning of what ``social media use'' actually means. Traditional metrics that document solely adoption counts, like the Reuters Digital News Report~\cite{reuters2025}, provide only a partial picture and may miss important practices that occur outside direct engagement. People may approach a platform to try it out, to selectively engage with it, and to later forget about it. As Jungselius and Weilenmann note, understanding social media use requires attention not only to the quantity of engagement but also to the quality and type of activity~\cite{Jungselius2018conceptualizing}. Social Media Journeys offer a lens for this: by capturing the full set of Transitions between platforms, we reveal how different generations structure their engagement over time, what practices they prioritize, and how they balance attention, well-being, and social connections. The interconnected nature of these transitions highlights patterns of movement, preference, and influence that linear or time-on-platform measures miss. 

\textbf{People Could Shape Social Media:}  
Our participants’ Journeys reveal that despite the dominance of central platforms--that are owned by a single company--people retain the agency to explore and seek out spaces that better align with their interests and realities. By distributing their communication channels, curating content they enjoy, and attending to their well-being, participants demonstrate a strong drive to navigate the social media landscape while staying anchored within a community. Historically, this general community has been centered on Facebook, yet many participants actively sought alternatives. We echo Kender and Spiel’s emphasis on creating customizable spaces and interest-centric~\cite{kenderBanalAutistic2023}, while also being community-oriented and safe~\cite{kender2025machine}. We encourage seeing social media as something that can be turned from a fixed environment into a space of user-driven possibility.

\textbf{Positioning Within Form–From:}
The Form--From framework \cite{zhangFormFrom2024} enabled us to align our self-reported data with an abstraction of social media technology from a content level. Even when the historical evolution of the design space could not be reproduced in our study, it underscores that being able to situate oneself in that framework is crucial. Even when new platforms and functions are situated in Threaded Commons, the appeal of other design spaces remains more relevant. We believe that the ``content'' in Form--From is what influences social media users the most, and positioning it well remains the challenge. While some platforms have content as their main attraction, platforms that are attractive for managing social ties would be expected to provide high-quality content either way. Future platforms must carefully consider what type of content they produce, where it is sourced from, and how it is presented.

\textbf{Reasons As Analytical Tools:}
Our findings show that reasons to transition to different platforms go beyond individual statements about likes or dislikes. They capture why social media matters in the first place, cutting across features to expose deeper themes of engagement such as trust, privacy, entertainment, and well-being. The reported reasons explain past platform Transitions and help anticipate future ones. They provide a shared vocabulary to compare movements across time, generations, and contexts. We therefore see our reported reasons as a key tool for future research, offering a way to map the dynamics of social media use and to understand its broader significance.

\textbf{Ask About Journeys:}
Our study shows that simple lists of platforms miss what actually matters about social media use. By applying our graph-based approach for eliciting context, we were able to understand preferences and motivations that adoption counts alone cannot capture. Use is not just about presence but about movements, comparisons, and strategies. Designers and researchers should therefore treat Social Media Journeys as narratives rather than inventories, foregrounding experience and significance over static measures of activity.

\subsection{Limitations and Future Work}

Methodologically, our approach foregrounds subjective meaning. Self-reports and retrospection on social media platforms are prone to nostalgic recollection and the participant's beliefs about a platform. Therefore, we treat Social Media Journeys as structured narratives that enable participants to reflect on their movements in a way that captures lived experience. This methodological lens offers a unique perspective on platform adoption, emphasizing experiential depth, temporality, and the relational context of movements across generations.

Finally, our study is situated in the United States. While most of the platforms discussed had their first substantial user base in the U.S., future work could apply similar methodologies in comparative, international contexts. Conducting this study in the U.S. allowed us to easily obtain a Census-matched quota-based sampling, which, for most countries, is not feasible. Extending the method across diverse sociotechnical environments would further enhance the generalization of cross-platform strategies, moderation practices, and user negotiation of well-being. 
\section{Conclusion}

In this paper, we introduced the concept of the \textit{Social Media Journey} to capture how people move between platforms and why. Using a quota-based U.S. sample and a novel graph-based tool, we mapped Journeys as networks of platforms and Transitions. Our results show that social media use is shaped not by isolated choices or single platforms, but by other platforms' influences, by community practices, and by broader societal pressures. Journeys reveal how people regulate participation, preserve relationships, and balance well-being in ways that differ across generations. Journeys show that adoption and migration are always shaped by prior experiences, competing alternatives, and broader social contexts. Any intervention or analysis must therefore situate platforms within these interdependent histories. By tracing Journeys across platforms and generations, we provide empirical and conceptual tools to inform the design, governance, and study of social media in ways that better align with lived realities.

\balance

\bibliographystyle{ACM-Reference-Format}
\bibliography{submission}
\end{document}